%% file: main.tex
\documentclass[acmsmall, screen]{acmart}

\AtBeginDocument{%
  \providecommand\BibTeX{{%
    Bib\TeX}}}

\usepackage[table]{xcolor}
\usepackage{tabularx}
\usepackage{adjustbox}
\usepackage{tcolorbox}
\usepackage{subcaption}
\usepackage{pifont}
\usepackage{enumitem}
\usepackage{makecell}
\usepackage[nameinlink]{cleveref}
\usepackage{multirow}
\usepackage{threeparttable}
\usepackage{xcolor}
\usepackage{listings}
\usepackage{afterpage}
\usepackage{placeins} 
\usepackage{float}
\usepackage{placeins} 
\usepackage{graphicx}
\usepackage{pgfplots}
\usepackage{pgfplotstable}
\usepackage{colortbl}
\usepackage{ifthen}
\definecolor{highcolor}{RGB}{204,255,204} 
\definecolor{lowcolor}{RGB}{255,204,204} 
\definecolor{codegreen}{rgb}{0,0.6,0}
\definecolor{codegray}{rgb}{0.5,0.5,0.5}
\definecolor{codepurple}{rgb}{0.58,0,0.82}
\definecolor{notered}{RGB}{200,0,0}
\definecolor{backcolour}{rgb}{0.95,0.95,0.92}

\lstdefinestyle{prompt}{
    commentstyle=\color{codegreen},
    keywordstyle=\color{magenta},
    stringstyle=\color{codepurple},
    basicstyle=\ttfamily\footnotesize,
    breakatwhitespace=false,         
    breaklines=true,                 
    captionpos=b,
    keepspaces=false,               
    showspaces=false,                
    showstringspaces=false,
    showtabs=false,                  
    tabsize=2,
    literate=
        {Ü}{{\"U}}1
        {ä}{{\"a}}1
        {ö}{{\"o}}1
        {ü}{{\"u}}1
        {≤}{{$\leq$}}1
        {→}{{$\rightarrow$}}1
        {-}{{-}}1
        {`}{{`}}1
        {—}{{---}}1
        {…}{{\ldots}}1
        {“}{{``}}1
        {”}{{''}}1
        {’}{{'}}1
        {‘}{{`}}1
        {"}{{"}}1
        {*}{{*}}1
        {<}{{<}}1
}

\lstdefinestyle{code}{
    backgroundcolor=\color{backcolour},   
    commentstyle=\color{codegreen},
    keywordstyle=\color{magenta},
    numberstyle=\tiny\color{codegray},
    stringstyle=\color{codepurple},
    basicstyle=\ttfamily\footnotesize,
    breakatwhitespace=false,         
    breaklines=true,                 
    captionpos=b,                    
    keepspaces=true,                 
    numbers=left,                    
    numbersep=5pt,                  
    showspaces=false,                
    showstringspaces=false,
    showtabs=false,                  
    tabsize=2
}

\pgfplotsset{compat=1.18}

\newboolean{showcomments}
\setboolean{showcomments}{true}
\ifthenelse{\boolean{showcomments}}
{\newcommand{\mynote}[2]{
\fbox{\bfseries\sffamily\scriptsize#1}
{\small$\blacktriangleright$\textsf{\emph{#2}}$\blacktriangleleft$}}}
{ \newcommand{\mynote}[2]{}}

\def\BibTeX{{\rm B\kern-.05em{\sc i\kern-.025em b}\kern-.08em
    T\kern-.1667em\lower.7ex\hbox{E}\kern-.125emX}}

\newcommand{\find}[1]{
\begin{tcolorbox}[leftrule=0.4mm,rightrule=0mm,toprule=0mm,bottomrule=0mm,left=0.0pt,right=0.0pt,top=0pt,bottom=0pt]
\em #1
\end{tcolorbox}
}

\title{Multi-Agent Strategies for Bridging Programming Language Gaps in Code Generation}

\begin{document}

\author{Micheline Bénédicte Moumoula}
\orcid{0009-0005-1206-2270}
\authornote{Corresponding author (micheline.moumoula@uni.lu)}
\affiliation{%
  \institution{University of Luxembourg }
  \city{Luxembourg}
  \country{Luxembourg}
}

\email{micheline.moumoula@uni.lu}

\author{Serge Lionel Nikiema}
\orcid{0009-0001-0066-3694}
\affiliation{%
  \institution{University of Luxembourg}
  \city{Luxembourg}
  \country{Luxembourg}
}
\email{lionel.nikiema@uni.lu}

\author{Albérick Euraste Djiré}
\orcid{0009-0007-0406-9573}
\affiliation{%
  \institution{University of Luxembourg}
  \city{Luxembourg}
  \country{Luxembourg}
}
\email{euraste.djire@uni.lu}

\author{Abdoul Kader Kaboré}
\orcid{0000-0002-3151-9433}
\affiliation{%
  \institution{University of Luxembourg}
  \city{Luxembourg}
  \country{Luxembourg}
}
\email{abdoulkader.kabore@uni.lu}

\author{Jacques Klein}
\orcid{0000-0003-4052-475X}
\affiliation{%
  \institution{University of Luxembourg}
  \city{Luxembourg}
  \country{Luxembourg}
}
\email{jacques.klein@uni.lu}

\author{Tegawendé F. Bissyandé}
\orcid{0000-0001-7270-9869}
\affiliation{%
  \institution{University of Luxembourg}
  \city{Luxembourg}
  \country{Luxembourg}
}
\email{tegawende.bissyande@uni.lu}

\renewcommand\shortauthors{Micheline Bénédicte MOUMOULA et al.}

\begin{abstract}
\input{sections/abstract}

\end{abstract}



\maketitle

\input{sections/introduction}
\input{sections/background_related_work}
\input{sections/approach}

\input{sections/evaluation}
\input{sections/results_analysis}
\input{sections/discussion}
\input{sections/conclusion}

\section*{Data Availability}
\label{sec:data_availability}
To promote transparency and facilitate reproducibility, we make our artefacts available to the community at: 
\begin{center}
\url{https://github.com/TruX-DTF/XL-CoGen}
\end{center}

\section*{Acknowledgments}
This work was supported 
by (1) the Luxembourg Ministry of Foreign and European Affairs through their Digital4Development (D4D) portfolio under project LuxWAyS, (2) the European Research Council (ERC) under the European Union’s Horizon 2020 research and innovation programme (Project NATURAL - grant agreement No.~949014), and (3) Google.org through the provision of Google Cloud Platform credits used to conduct our experiments.

\bibliographystyle{ACM-Reference-Format}
\bibliography{acmart}

\end{document}

%% file: sections/abstract.tex
Real-world software systems are inherently multilingual, but current large language models are not equally consistent across programming languages. This mismatch limits code generation usefulness, especially for underrepresented languages. Existing approaches improve code generation through fine-tuning, multi-agent reasoning, or translation, but remain language-specific, assume existing source code, or rely on per-language test suites.

We introduce XL-CoGen, a three-stage multilingual code-generation pipeline that starts from a natural-language specification and a shared test list to generate correct implementations across multiple target languages.
XL-CoGen first validates direct generation by constructing and correcting the test harness; when it fails, it transfers through empirically selected intermediate languages and translates validated solutions; it then repairs the best candidate through diagnosis and minimal patching. This design requires neither target-language training nor language-specific test suites. Across two benchmarks and multiple LLMs, XL-CoGen consistently improves over direct generation, with the largest gains on low-performing languages. In our Rust fine-tuning case study, XL-CoGen outperforms the best fine-tuned baseline by 22 percentage points and improves challenging languages by up to 33 points on multilingual benchmarks.

Ablation results show that transfer and repair are complementary: repair suffices on easier tasks, whereas transfer becomes more important as difficulty increases, especially for weak target languages.

%% file: sections/introduction.tex
\section{Introduction}
\label{sec:introduction}
Large Language Models (LLMs) have evolved far beyond their original text generation capabilities, emerging as powerful computational tools that achieve strong performance across diverse technical domains, including language understanding, formal reasoning, and program synthesis~\cite{kaddour2023challenges, gao2023palprogramaidedlanguagemodels, zheng2024surveylargelanguagemodels, kaddour2023challengesapplicationslargelanguage}. Code-specialized and general-purpose foundation models~\cite{anil2023palm, awais2025foundation, roziere2023code, zhu2024deepseek, qwen2.5, wei2024selfcodealign} now support a wide range of software engineering tasks---including code completion, synthesis, translation, and repair~\cite{chen2021evaluating, roziere2023code, chen2025surveyevaluatinglargelanguage}. Their ability to process and generate code across multiple programming languages has made them increasingly valuable tools in heterogeneous software development environments~\cite{barke2023grounded, bird2022taking, mishra2024granitecodemodelsfamily}.

Despite these advances, significant challenges remain for \emph{cross-language code generation}. Empirical evaluations consistently show that LLM performance varies substantially across programming languages. High-resource languages such as Python, JavaScript, and Java typically exhibit much higher code generation success rates than lower-resource languages such as Julia, Rust, Perl, or Haskell~\cite{orlanski2023measuringimpactprogramminglanguage, peng2024humanevalxlmultilingualcodegeneration, anil2023palm2technicalreport}. These disparities arise from multiple factors, including imbalanced training data, differences in language ecosystems, and varying levels of syntactic and semantic complexity~\cite{cassano2024knowledgetransferhighresourcelowresource, macedo2024intertransleveragingtransitiveintermediate}. This matters in practice because modern software systems are rarely monolingual: different components often target different platforms, execution environments, or ecosystem constraints, leading teams to implement related functionality across multiple programming languages. As LLMs are increasingly used as software-development assistants, their practical value depends not only on strong performance in popular languages such as Python, but also on reliable generation across the broader multilingual environments found in real systems.

Recent benchmarks further highlight these limitations. Cross-language evaluation datasets such as CrossCodeEval~\cite{ding2023crosscodeevaldiversemultilingualbenchmark} expose the difficulty of generating correct code across diverse programming languages, particularly when tasks involve non-trivial reasoning or cross-file context. At the same time, recent work on cross-language translation and repair has shown that performance asymmetries between languages can sometimes be exploited to improve outcomes in low-resource languages~\cite{luo2025unlockingllmrepaircapabilities}. However, existing approaches typically address only partial aspects of the problem. Many frameworks focus on single-language generation, while translation-based methods assume the presence of an existing program rather than generating code directly from a natural-language specification.

Recent multi-agent code generation systems such as 
CodeCoT~\cite{huang2024codecottacklingcodesyntax}, AgentCoder~\cite{huang2024agentcodermultiagentbasedcodegeneration}, 
LDB~\cite{zhong2024debuglikehumanlarge}, MapCoder~\cite{islam2024mapcodermultiagentcodegeneration}, and CodeCoR~\cite{pan2025codecorllmbasedselfreflectivemultiagent} demonstrate that iterative reasoning, planning, and self-reflection can significantly improve code generation quality. These systems coordinate multiple specialised agents to analyse requirements, generate code, test candidate implementations, and refine solutions. However, they are predominantly evaluated in single-language environments—most often Python—and therefore do not directly address the systematic performance disparities observed across programming languages in multilingual benchmarks.

More broadly, many current frameworks and evaluation settings implicitly privilege well-supported languages. They improve code generation within strong-language settings, but leave open the practical question that matters in multilingual software engineering: how to systematically recover performance for weaker target languages from the same specification, without requiring language-specific training data or separate handcrafted evaluation pipelines for each language.

Recent work has begun to exploit performance asymmetries across programming languages to improve LLM-based software engineering tasks. For example, LANTERN~\cite{luo2025unlockingllmrepaircapabilities} leverages cross-language translation to improve automated program repair by translating buggy programs from languages where repair performance is weak to languages where models demonstrate stronger repair capabilities, before refining the patch through multi-agent interaction. Similarly, INTERTRANS~\cite{macedo2024intertransleveragingtransitiveintermediate} improves code-to-code translation by exploring multi-hop translation paths through intermediate programming languages, decomposing large syntactic and semantic gaps into smaller translation steps. Mora et al.~\cite{mora2024synthetic} propose Synthetic Programming Elicitation and Compilation (SPEAC), which generates programs in a dedicated intermediate representation that is subsequently compiled into very low-resource programming or formal languages.

While these approaches demonstrate that cross-language transfer can improve translation, repair, or low-resource generation, they operate under assumptions that differ from the setting of multilingual text-to-code generation. Repair and translation approaches assume the existence of an initial program that can serve as a starting point for cross-language transfer, while SPEAC introduces a specialised intermediate representation and compilation pipeline tailored to specific target languages. In contrast, multilingual text-to-code generation starts from a natural-language specification without any initial program and must ensure correctness across multiple target languages under a shared behavioural specification.

This setting introduces two challenges. First, the system must reliably determine when a generated implementation is correct before using it as a basis for cross-language transfer. Second, the generation process must remain consistent across languages without relying on language-specific test suites or specialised intermediate representations.

Our approach addresses these challenges by designing a pipeline in which correctness validation precedes cross-language transfer and where a single behavioural specification drives generation across languages. Rather than translating arbitrary generated programs or introducing a dedicated intermediate representation, we search for correct implementations in existing programming languages where the model demonstrates strong generation capabilities. Candidate implementations are validated against a fixed set of behavioural tests derived from the task specification, and only validated solutions are transferred across languages.

Based on this principle, we propose XL-CoGen, a three-stage pipeline for multilingual code generation driven by a single natural-language specification and a shared test list. The pipeline first establishes a reliable evaluation interface by generating language-specific harnesses and validating direct code generation in the target language. When direct generation fails, the system attempts synthesis in empirically selected intermediate languages and translates the solution to the target language only after it passes all tests in the intermediate environment. Finally, the pipeline repairs the best remaining candidate using structured diagnosis and minimal patching. Throughout the process, artefacts such as function signatures, harnesses, and validated intermediate implementations are reused across stages and languages.

We evaluate XL-CoGen on two widely used programming benchmarks (MBPP and LeetCode), spanning 17 programming languages and multiple LLMs. Our experiments show that systematically leveraging cross-language transfer can substantially improve generation success, particularly for languages where direct generation performs poorly.

Our work makes the following key contributions:

\begin{itemize}[leftmargin=*]

\item \textbf{Data-driven intermediate language selection.}
We introduce a data-driven procedure to select intermediate languages from precomputed cross-language transfer matrices that combine generation success in a candidate language and translation success from that language to the target. We show that LLM-based selection often favours high-resource languages (e.g., Python) even when they are suboptimal for translating into a given target. Our transfer-matrix-based rankings instead use observed translation compatibility, yielding more effective Stage~2 recovery. Ablation studies confirm that this selection strategy significantly outperforms LLM-based choice across target languages, with pronounced gains for lower-resource languages such as Erlang, OCaml, and Perl.

\item \textbf{XL-CoGen multilingual generation pipeline.}
We present a three-stage pipeline that (i) validates direct generation in the target language and fixes evaluation setup via harness generation and correction, (ii) when direct generation fails, generates solutions in empirically chosen intermediate languages and translates to the target only when the intermediate solution passes all tests, and (iii) repairs the best failing candidate through diagnosis and minimal patching. Unlike prior work that applies uniform treatment across languages or costly runtime selection, we reuse artefacts such as signatures, harnesses, and validated intermediate solutions across stages, so each stage builds on the previous one without discarding intermediate results. The methodology targets natural language-to-code generation from a single specification, where no initial code exists, in contrast to repair-oriented approaches that assume existing buggy implementations.

\item \textbf{Performance compared to prior work.} Evaluation on two benchmarks (MBPP and LeetCode), 17 programming languages, and multiple state-of-the-art LLMs shows that XL-CoGen consistently improves over direct generation, with the largest gains on initially weak target languages. On a controlled Rust fine-tuning case study using the best 
available Rust-specific dataset, XL-CoGen outperforms the best 
fine-tuned baseline by 22 percentage points. It also improves performance on challenging languages by up to 33 points, depending on the benchmark and model. On the selected languages used for comparison with prior tools, XL-CoGen also outperforms the other tools in all reported Gemini-2.5-pro cases.
\end{itemize}

\begin{figure*}[!ht]
        \centering
        \includegraphics[width=\linewidth]{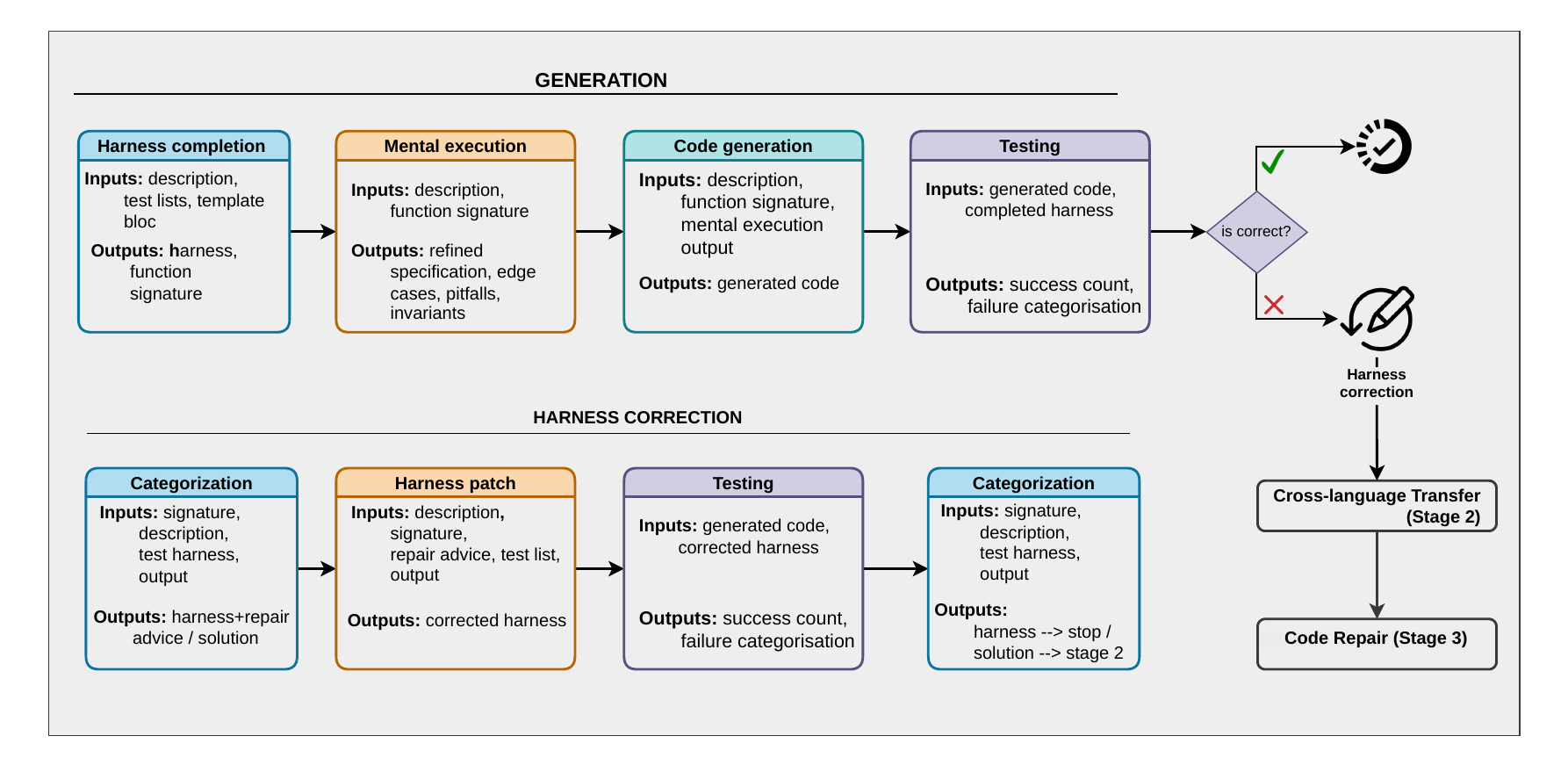} 
        \vspace{-0.5cm}
    \caption{Overview of the XL-CoGen workflow}
    \label{fig:workflow}
\end{figure*}

%% file: sections/background_related_work.tex
\section{Background and Related Work}
\label{sec:background}

This section provides an overview of the relevant literature and foundational work that informs our approach to cross-language code generation. We begin by examining existing benchmarks and the documented performance disparities across programming languages that motivate our work. We then review recent advances in multi-agent code generation systems, which primarily focus on single-language environments, followed by an analysis of cross-language translation and repair techniques. Finally, we position our contribution within this landscape, highlighting the gaps our framework addresses in multilingual code generation.

\subsection{Cross-Language Code Generation Benchmarks and Performance Disparities}

Multiple benchmarks have been developed to evaluate code generation across programming languages, including HumanEval-X~\cite{zheng2023codegeex}, HumanEval-XL~\cite{peng2024humanevalxlmultilingualcodegeneration}, MultiPL-E~\cite{10103177}, ~\cite{xu2025cruxevalxbenchmarkmultilingualcode}, MBXP~\cite{athiwaratkun2023multilingualevaluationcodegeneration}, CrossCodeEval~\cite{ding2023crosscodeevaldiversemultilingualbenchmark}, and BabelCode~\cite{roziere2020unsupervised}. 

Empirical evaluation on these benchmarks consistently reveals significant performance variations across programming languages. Languages with abundant training data representation, such as Python, Java, and JavaScript, achieve substantially higher performance compared to lower-resource languages like Rust, Julia, Haskell, and Perl. These disparities reflect both training data imbalances and the inherent limitations of current evaluation frameworks in capturing the full complexity of multilingual code generation. \Cref{fig:motivation} illustrates these performance gaps using GPT-3.5-Turbo and GPT-4.1-Mini results on the BabelCode-MBPP.

\subsection{Multi-Agent Code Generation Systems}
Recent advances in multi-agent architectures and execution-guided feedback loops have demonstrated promising improvements in code generation quality, though existing frameworks predominantly target single-language settings, with a heavy focus on Python development. 

Several approaches introduce dedicated iterative paradigms that generate or utilise test execution to guide code construction. CodeCoT~\cite{huang2024codecottacklingcodesyntax} integrates chain-of-thought reasoning with self-generated test cases, prompting the model to produce both a candidate solution and a set of verification tests in a single pass before iterating based on execution 
feedback. Rather than relying on whole-program feedback, LDB~\cite{zhong2024debuglikehumanlarge} takes a finer-grained approach to code repair by segmenting program execution into sequential basic blocks, enabling the LLM to inspect intermediate runtime states and isolate faults segment by segment. While effective, these systems remain tightly coupled to the evolving, model-generated solutions and assume direct access to a stable, language-specific executable runtime.

A distinct class of frameworks addresses code generation through multi-agent coordination with explicit, specialised roles. AgentCoder~\cite{huang2024agentcodermultiagentbasedcodegeneration} introduced a multi-agent framework with specialised roles for code generation, testing, and execution-based feedback, enabling iterative code generation and validation through feedback-driven refinement. MapCoder~\cite{islam2024mapcodermultiagentcodegeneration} proposed a multi-agent planning and execution workflow where agents specialise in distinct aspects of the development lifecycle, including requirement analysis, algorithm design, implementation, and verification. CodeCoR~\cite{pan2025codecorllmbasedselfreflectivemultiagent} developed a collaborative framework enabling multiple specialised agents to generate, review, and refine code through multi-turn, self-reflective interactions.

Despite their success in single-language environments, these multi-agent and iterative approaches have not been systematically extended to cross-language scenarios. Their architectures are fundamentally designed around homogeneous language ecosystems, creating a significant gap for frameworks that can effectively orchestrate generation, validation, and artefact reuse across diverse programming language ecosystems.

\subsection{Cross-Language Translation and Repair}

Cross-language translation techniques have emerged as a promising approach for addressing performance disparities in multilingual code generation. Luo et al. introduced LANTERN, a framework addressing code repair in low-resource programming languages through cross-language translation and multi-agent refinement. LANTERN translates buggy code from low-resource to high-resource languages, applies LLM-based repair techniques, and translates the corrected code back to the original language. While effective in exploiting performance asymmetries between languages, this approach focuses primarily on code repair rather than generation from scratch and may introduce semantic drift through multiple translation steps.

Macedo et al. developed INTERTRANS, which leverages transitive intermediate translations to enhance LLM-based code translation. Rather than attempting direct source-to-target translation, INTERTRANS employs a Tree of Code Translation (ToCT) algorithm to plan multi-hop translation sequences through intermediate programming languages. This methodology decomposes complex syntactic and semantic gaps into smaller, more manageable translation steps. Evaluated across CodeNet, HumanEval-X, and TransCoder benchmarks, INTERTRANS achieved 18.3\% to 43.3\% absolute improvements over direct translation methods.

Mora et al.~\cite{mora2024synthetic} propose Synthetic Programming Elicitation and Compilation (SPEAC), a framework for text-to-code generation in very low-resource programming and formal languages (VLPLs). Rather than generating code directly in the target language $T$, SPEAC identifies a parent language $P$ that aligns with the model's natural generation tendencies, together with a constrained child language $C \subset P$ that admits a formal compiler $f : C \to T$. A two-stage repair pipeline then brings generated programs into $C$. The first stage computes a maximally consistent subtree of the generated AST via a MAX-SMT formulation, replacing constraint-violating nodes with holes. The second stage uses model-driven repair to resolve remaining holes, with unresolved holes iteratively delegated back to the LLM. While this approach yields substantial improvements in syntactic correctness over few-shot, self-repair, and fine-tuning baselines, the selection of $P$ and $C$ relies on a structured but manual elicitation process that remains language-specific and labour-intensive, limiting portability to broader multilingual programming settings. By contrast, XL-CoGen targets standard programming languages and replaces the manual elicitation step with a data-driven merit score that jointly captures model reliability and cross-language transfer success, enabling systematic intermediate language selection across diverse benchmarks.

\subsection{Positioning of Our Work}

We position our work by analysing limitations of existing approaches to LLM-based code generation and by clarifying how our pipeline addresses them.

\textbf{Test generation from code.}  
Several recent frameworks rely on generating tests during the coding process and iterating until the produced code passes those tests. Approaches such as CodeCoT follow this paradigm, generating candidate tests for the program under development. More generally, frameworks that derive tests from evolving model-generated code are exposed to a \emph{solution--test misalignment} risk: when the generated code is wrong, the generated tests may become aligned with that incorrect behaviour. In such cases, the code and the tests can appear mutually consistent while both violate the intended problem specification. This risk is especially relevant when the same model is responsible for both interpreting the natural-language requirement and producing the artefacts used for validation.  
CodeCoR attempts to improve test quality through multi-agent collaboration and a dedicated Test Agent, but the generated tests are still tied to the evolving candidate solution, which does not fully eliminate the possibility of aligned errors or insufficient edge-case coverage.  

\noindent
Our framework avoids this pitfall by treating the problem specification and the expected behaviour as \emph{fixed inputs}. The system begins with a single natural-language description and a predefined list of tests (e.g., input/output pairs or assertions as provided in benchmarks such as MBPP). Tests are not derived from the generated code. Instead, they serve as the authoritative specification that drives harness generation, validation, and comparison across all target languages.

\textbf{Dependency on language-specific test suites.}  
Many multi-agent systems, such as MapCoder, are evaluated in settings where executable test suites are already available for the target programming language. In multilingual development scenarios, reproducing such a setting typically requires maintaining separate test suites for each language implementation, which is costly and often impractical.  

\noindent
In contrast, our framework starts from a \emph{single source of truth}: one natural-language specification and one test list. The test specification is automatically adapted to each target language by generating language-specific harnesses and by translating assertion syntax and runtime structures when needed. This allows the same behavioural specification to drive evaluation across multiple programming languages without requiring pre-existing test suites for each language.

\textbf{Single-language focus of multi-agent code generation.}  
Frameworks such as CodeCoT, AgentCoder, LDB, MapCoder, and CodeCoR demonstrate that coordinated multi-agent reasoning can improve code generation quality. However, these systems are primarily evaluated in single-language environments, most often Python, and do not explicitly address the performance disparities across programming languages documented in multilingual benchmarks. Their architectures are not designed to orchestrate generation, evaluation, and artefact reuse across multiple languages from a shared specification.

\textbf{Translation- and representation-based approaches.}  
Several recent works attempt to address cross-language challenges through translation or specialised representations. LANTERN~\cite{luo2025unlockingllmrepaircapabilities} focuses on repairing existing buggy programs by translating them into a high-resource language, performing repair, and translating the corrected code back to the original language. While effective for repair, this approach assumes the existence of an initial program and may introduce semantic drift during round-trip translation.  
INTERTRANS~\cite{macedo2024intertransleveragingtransitiveintermediate} improves code-to-code translation by exploring multi-hop translation paths through intermediate languages, but it assumes the availability of a correct source program and does not address generation from natural-language specifications.  
SPEAC~\cite{mora2024synthetic}, as discussed above, improves generation for extremely low-resource languages, but relies on designing dedicated intermediate languages and compilation pipelines tailored to specific targets.

\noindent
Taken together, prior work addresses important adjacent settings---single-language code generation, test-guided refinement, code-to-code translation, or repair from an existing program. What remains insufficiently addressed is multilingual \emph{text-to-code} generation from a single natural-language specification under a shared behavioural specification, where no initial source program is available and where recovery must explicitly handle uneven model performance across target languages.

We address a complementary problem: \emph{multilingual code generation directly from a natural-language specification}. Our framework starts from a single description and a single list of behavioural tests and orchestrates generation across languages through a structured pipeline.  

First, the system constructs a reliable evaluation interface by generating language-specific harnesses and structured specifications before producing code. Second, when direct generation in the target language fails, the system generates candidate solutions in an intermediate language and translates them to the target only after they successfully pass all tests in the intermediate environment. Third, the framework attempts to repair the best available failing solution using generic diagnostic guidance while keeping the test cases hidden from the model prompt to reduce overfitting.

This design enables a unified multilingual generation workflow where a single specification drives code synthesis, validation, translation, and repair across languages. Prior work has advanced multi-agent code generation, cross-language repair, and generation via intermediate representations, but remains constrained by reliance on code-derived tests, language-specific test suites, existing source programs, or dedicated intermediate representations. Our framework builds on these advances by leveraging pipeline design, artefact reuse, and controlled intermediate-language transfer to support multilingual code generation from a single natural-language specification without those dependencies.

%% file: sections/approach.tex
\section{Methodology}
\label{sec:methodology}
We present XL-CoGen, a structured three-stage methodology for multilingual code generation that improves performance across programming languages through strategic cross-language knowledge transfer. Illustrated in~\Cref{fig:workflow}, our approach is built on the core insight that large language models exhibit non-uniform generation quality across languages, and that this performance asymmetry can be systematically leveraged to improve generation in underperforming target languages by transferring through more reliable intermediate languages.

\begin{figure*}[!ht]
        \centering
        \includegraphics[width=\linewidth]{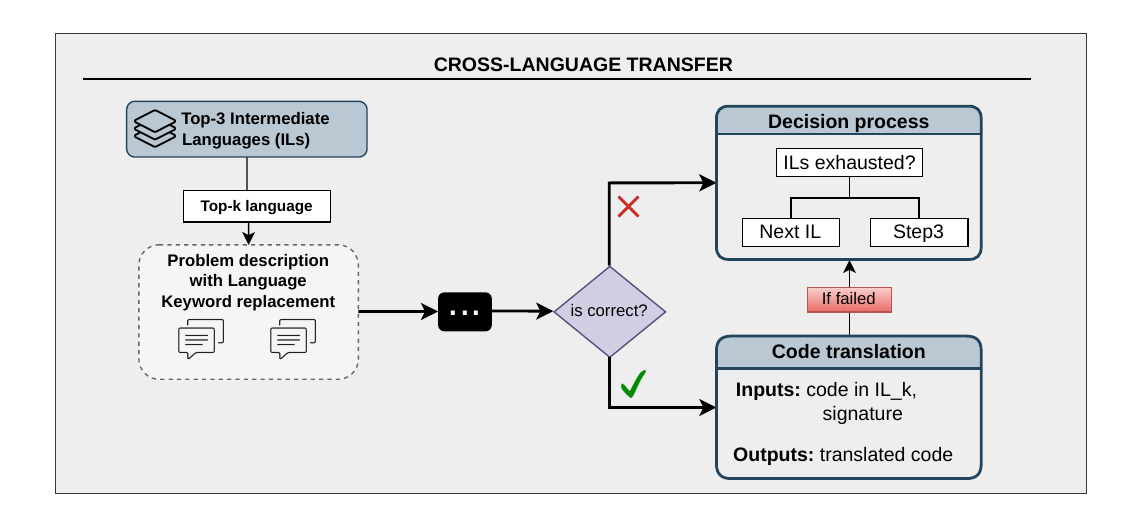} 
        \vspace{-0.5cm}
    \caption{Workflow of the cross-language multi-agent transfer process}
    \label{fig:workflow2}
\end{figure*}

\begin{figure}[!ht]
        \centering
        \includegraphics[width=\linewidth]{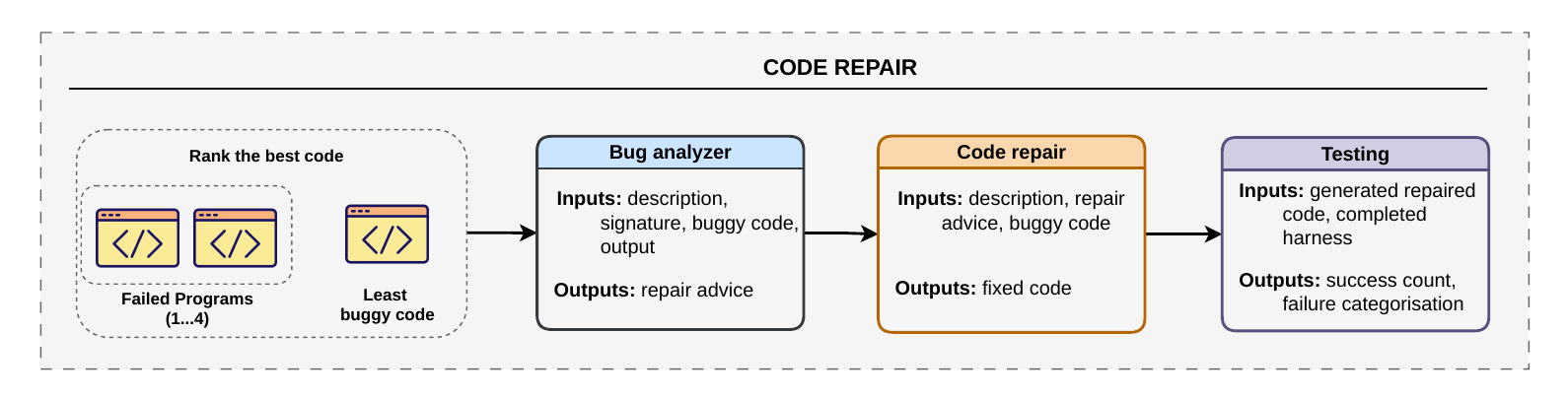} 
        \vspace{-0.5cm}
    \caption{Workflow of the multi-agent repair process}
    \label{fig:workflow3}
\end{figure}

The framework operates on a single natural-language problem specification and a fixed test list, which together define a shared behavioural specification for generating correct implementations across multiple target languages. It progressively applies validated direct generation, cross-language transfer, and refinement while reusing artefacts across languages to preserve semantic fidelity throughout the process. \textit{Stage 1} establishes a direct solution in the target language by generating and correcting the test harness, then generating code to satisfy the extracted signature and behavioural specification. \textit{Stage 2 (~\Cref{fig:workflow2})} leverages more reliable languages when direct generation fails, translating successful solutions from intermediate languages into the target. \textit{Stage 3 (~\Cref{fig:workflow3})} refines remaining failures through diagnosis and minimal patching rather than full re-generation. Each stage has a well-defined role and addresses a distinct source of failure---evaluation setup, language asymmetry, or solution logic---so that the pipeline tackles the right problem at the right time.

\subsection{Stage 1: Harness Generation, Correction, and Code Generation}
\label{sec:stage1}
Stage 1 produces the first solution in the target language and resolves evaluation-setup issues before any cross-language transfer. Naive generation from a natural-language problem alone leads to interface mismatches and misalignment between the test harness and the generated code. We address this by generating the test harness and a precise function specification first, then generating code that conforms to that specification and the harness interface, and by diagnosing and repairing harness faults when tests fail, so that correct solution code is not altered unnecessarily. The tests themselves are treated as fixed behavioural inputs and are never generated from candidate code.

Stage 1 comprises four components.

\textbf{Harness generation and signature extraction.} From the problem statement and the given test list, we generate a complete, language-specific test harness using language-specific templates. The harness embeds the \emph{function signature}: a short natural-language comment and the exact function declaration that the solution must implement. We extract this signature and reuse it in every subsequent step. Doing so ensures that all generated code conforms to the same interface expected by the tests, which eliminates signature drift and many compilation and linking failures. The harness also provides a uniform execution and validation layer across all languages.

\textbf{Mental execution.} Before generating code, the model reasons about the problem given the natural-language problem description and the extracted signature. It produces a trace of expected behaviour, edge cases, pitfalls, and invariants, without writing any code. This yields a concise, structured specification that the code generator follows, reducing ambiguity and providing a clear target for implementation. It produces a specification of intended behaviour prior to the implementation step.

\textbf{Code generation and validation.} Using the problem, the extracted signature, and the mental-execution specification, we generate the solution code in a single step. The code is merged with the pre-generated harness, compiled, and executed in our test environment. The outcome is either full pass, partial pass, or failure (e.g.\ build error, syntax error, or runtime error). When all tests pass, Stage 1 terminates for that entry. Otherwise we apply harness correction.

\textbf{Harness correction.} A large fraction of early failures stems from the test harness (e.g., incorrect driver calls, signature mismatches, or language-specific formatting) rather than from the solution logic. We therefore apply a two-step procedure. \emph{Diagnosis:} the model receives the \emph{full executable template} (solution plus harness) and the compiler or runtime output; from the error location and structure it returns a \emph{categorization}---whether the fault lies in the harness or in the solution---along with structured advice when the fault is in the harness. \emph{Patching:} when the fault is attributed to the harness, a second model call receives \emph{only} the buggy harness (not the solution) plus the signature, the correction advice, and the test list; it outputs the corrected harness only. The solution code is thus never sent in the patching step and remains unchanged. We then re-run the tests. We consider the harness \emph{validated} when it is deemed reliable (fault was in the solution, or the harness was successfully patched and tests pass or re-verification attributes remaining failures to the solution); the harness is considered \emph{unfixable} when the fault was attributed to it but no valid corrected harness could be obtained or re-verification still points to the harness. Only entries with a validated harness are passed to Stage 2 and Stage 3, so that transfer and code correction operate on a trusted evaluation scaffold. This step ensures that we fix the evaluation setup before invoking transfer or solution-level correction and recovers solutions that were correct but were previously judged failing due to a faulty test driver.

Stage 1 thus delivers a strong baseline. Entries that still fail are attributable to solution logic or to limitations of direct generation in the target language, which Stage 2 and Stage 3 address.

\subsection{Stage 2: Transfer via Intermediate Languages}
\label{sec:stage2}

LLMs are not equally reliable in all languages. When direct generation in the target language $l_t$ fails at the end of Stage 1, we transfer knowledge from languages where the model is more reliable. We generate a solution in an \emph{intermediate language} $l_i$ using the same Stage 1 process with the language parameter set to $l_i$ instead of $l_t$---i.e.\ we run the Stage 1 pipeline in $l_i$, including harness validation when needed. We translate to $l_t$ only when that solution achieves full success in $l_i$. This reframes ``hard'' target-language generation as ``easier'' generation in $l_i$ plus translation, turning language asymmetry into a lever for improvement. Stage 2 is applied only to entries that still fail after Stage 1 and have a validated harness.

\textbf{Intermediate language selection.} \emph{Unlike IR-based approaches such as SPEAC, which fix a single manually-designed intermediate representation per target language, XL-CoGen selects intermediate languages empirically and adaptively.} For each target language $l_t$, we rank candidate intermediate languages using a data-driven merit score $M(l_i \mid l_t) = R(l_i)^\alpha \cdot T(l_i \to l_t)^\beta$, where $R(l_i)$ is the standalone generation success rate in language $l_i$ and $T(l_i \to l_t)$ is the observed success rate of translating correct solutions from $l_i$ to $l_t$. We select the top-3 intermediate languages per target. This ensures that we use languages where the model is both likely to produce a correct solution and likely to translate it successfully.

\textbf{Generation in $l_i$ and translation.} For each failing entry, we take the top-3 intermediate languages in order. For each candidate $l_i$, we run the Stage 1 generation process with the language parameter set to $l_i$ (the prompt and harness generation use $l_i$ instead of the target language). We obtain a solution in $l_i$ and run the tests in $l_i$. Only when the solution achieves full success in $l_i$ do we translate it to the target language; the model preserves behaviour and respects the function signature. The translated code is merged with the existing harness (built for $l_t$), and the tests are run. We try intermediate languages in order until one yields a full pass in $l_t$. This step recovers tasks that Stage 1 could not solve in $l_t$, especially when the model is weaker in the target language, without requiring any additional target-language training data.

Stage 2 is the main lever for improving performance in underperforming languages without requiring additional target-language training data.

\subsection{Stage 3: Code Correction}
\label{sec:stage3}

Stage 3 repairs the best available failing solution from previous stages through systematic error diagnosis and targeted correction. Its input is the union of outcomes from Stage 1 (direct generation in the target language) and Stage 2 (translated solutions from intermediate languages). For each entry that still fails, we have one or more candidate solutions---the solution from Stage~1 and, if Stage~2 was run, one or more translated solutions (one per intermediate language tried). \textit{Best-solution selection:} We rank these candidates by a score that favours successful build or run, then the number of passing tests, and we select the best candidate as the starting point for correction (i.e.\ the one closest to passing). We apply a two-step repair: first, diagnose the failure in natural language (cause and fix), then apply that advice to produce a minimal patch.

Stage 3 correction consists of two steps, applied iteratively (up to three attempts) when tests still fail.

\textbf{Repair advice (diagnosis).} The model receives the problem description, the function signature, the full template that was executed, and the execution output. It produces a short, structured \emph{repair advice} in two parts: the \emph{cause} (what is wrong, in generic terms, without citing specific test values) and the \emph{fix} (what change to make, specific enough to be applied in one edit). We exclude raw test cases and concrete input-output values from this prompt so that the advice generalizes. This step converts execution feedback into an actionable, human-interpretable instruction and reduces the risk of the patching step making unrelated changes.

\textbf{Code patching.} A second model calls receives the problem, the signature, the repair advice, and the buggy solution code. The model outputs only the corrected solution code. We merge it with the harness and re-run the tests. This step applies the minimal necessary change and avoids full re-generation, improving both success rate and efficiency.

Stage 3 thus provides a final, targeted recovery step that complements transfer: it addresses solution-level faults, whether they come from direct generation (Stage 1) or from translated code (Stage 2), keeping the pipeline efficient and focused.

%% file: sections/evaluation.tex
\section{Evaluation}
\label{sec:experimental_setup}

This section outlines our experimental methodology for evaluating XL-CoGen's cross-language code generation capabilities. We begin by articulating our research questions, describe the models and datasets used in our study, define our evaluation metrics, and conclude with our experimental setup.

\subsection{Research Questions} 
\label{rqs}

Our evaluation is guided by three research questions that systematically assess the effectiveness of our cross-language code generation framework:

\begin{itemize}[leftmargin=*]
    \item \textbf{RQ1}: \emph{How effective is XL-CoGen across programming languages and models?} We measure how well the pipeline generates correct code across target languages, including underperforming languages where direct generation is difficult.
    \item \textbf{RQ2}: \emph{How does XL-CoGen compare with prior approaches, including language-specific fine-tuning and existing code-generation tools?} We compare XL-CoGen against relevant baselines and prior systems to assess whether its gains come from the pipeline design itself rather than from target-language specialisation.
    \item \textbf{RQ3}: {\em What is the impact of individual components of XL-CoGen on its overall performance?} This question examines each stage's contribution within the XL-CoGen pipeline to identify the most critical components for effective multilingual code generation.
\end{itemize}

\subsection{Models}
\label{lab:models}
We systematically selected three strong code-capable LLMs to cover different providers, access models, and capability profiles while keeping the evaluation relevant to current practice. Specifically, we include one widely used proprietary model from OpenAI, one strong open-weight model from DeepSeek-V3.2, and one competitive proprietary model from Google. This design broadens the evaluation across both deployment settings and model families, so that the empirical conclusions do not rest on a single provider, a single model lineage, or a single style of coding performance.

\textbf{OpenAI proprietary model.} We selected GPT-4.1-mini as a widely used proprietary model with strong coding performance and a practical latency-cost profile for agent-based workflows~\cite{openai2024gpt4technicalreport}. Results from the EvalPlus leaderboard~\cite{evalplus} also indicate that GPT-4-class models perform strongly on HumanEval+ and MBPP+~\cite{geminiteam2024gemini15unlockingmultimodal}. This makes GPT-4.1-mini a relevant proprietary reference point for code generation and repair in our setting.

\textbf{DeepSeek open-weight model.} We selected DeepSeek-V3.2~\cite{deepseekai2024deepseekv3technicalreport} as our open-weight representative because it is a strong current model for code-related tasks. This inclusion ensures that our findings are not limited to proprietary systems and provides a competitive open-weight reference point for multilingual code generation.

\textbf{Google proprietary model.} We selected Gemini-2.5-pro to broaden the evaluation across providers and capability profiles. Gemini models have been reported to perform strongly on reasoning- and coding-oriented evaluations~\cite{geminiteam2024gemini15unlockingmultimodal}, which is relevant for XL-CoGen because the pipeline relies not only on code generation but also on structured diagnosis, validation, translation fidelity, and targeted repair. Including Gemini lets us test whether the gains of XL-CoGen hold under a third provider and a distinct model profile rather than reflecting idiosyncrasies of only GPT-style or DeepSeek-style systems.

For all experiments, we used greedy decoding with a temperature of 0 to mitigate randomness while preserving output coherence, following established practices in the code generation literature~\cite{chen2021evaluating}.

\subsection{Datasets}
\label{lab:datasets}

We evaluate XL-CoGen on two complementary benchmarks that capture different aspects of multilingual code generation.

\textbf{MBPP benchmark.} Our primary benchmark follows the task format of MBPP (Mostly Basic Python Problems)~\cite{austin2021program}: each instance consists of a natural-language problem description together with a test list expressed as assertions or input/output examples. Our MBPP evaluation uses 427 problems. This benchmark is particularly suitable for our setting because it starts from a single natural-language specification and an executable test list, matching the inputs required by XL-CoGen. It also provides a controlled environment for analysing stage-wise improvements and intermediate-language selection. To support multilingual evaluation, we adapt the problem statements to the target language: when a description explicitly mentions Python, we replace it with the target language name; when no language is specified, we append a short clarification indicating the required target language. This yields a uniform MBPP-style benchmark across languages while preserving the original task semantics and test behaviour.

\textbf{LeetCode benchmark.} Our second benchmark follows the task format of LeetCode~\cite{xia2025leetcodedatasettemporaldatasetrobust}: each instance consists of a natural-language problem description, starter code, and an executable set of tests. Our LeetCode evaluation uses 228 problems. This benchmark is particularly suitable for our setting because it complements MBPP tasks with more diverse and generally more challenging algorithmic problems, thereby stressing reasoning, edge-case handling, and robustness on problems with less formulaic solution patterns. To construct multilingual LeetCode instances, we retain the original problem statement as the prompt and extract the target function name from the starter code. We then derive a compact executable test list from the original assertion-based tests by first identifying public examples already present in the problem statement, then sampling up to five assertions while excluding, whenever possible, tests that duplicate those public examples. The sampled assertions are rewritten to replace the generic \texttt{candidate(...)} call with the recovered function name. For multilingual evaluation, each task is instantiated for each of the 17 target languages by appending an explicit instruction requesting that the problem be solved in that language. This yields a multilingual LeetCode benchmark aligned with the same evaluation protocol as our MBPP tasks while preserving the original problem semantics and executable behaviour.

Across the two benchmarks, our experiments cover 17 programming languages in total.

\textbf{Translation dataset for intermediate-language selection.} To estimate the translation component used in Stage~2, we additionally use a translation benchmark derived from MultiPL-E~\cite{cassano2022multiplescalableextensibleapproach}. We select 50 problems and evaluate translation across 16 programming languages. This yields 12{,}000 directed source-to-target translation instances in total (50 problems $\times$ 16 source languages $\times$ 15 target languages). These translation results are used only to estimate cross-language transfer reliability for intermediate-language selection; they are not part of the main MBPP or LeetCode generation evaluation.

\begin{figure}[!ht]
         \centering
         \includegraphics[width=0.6\linewidth]{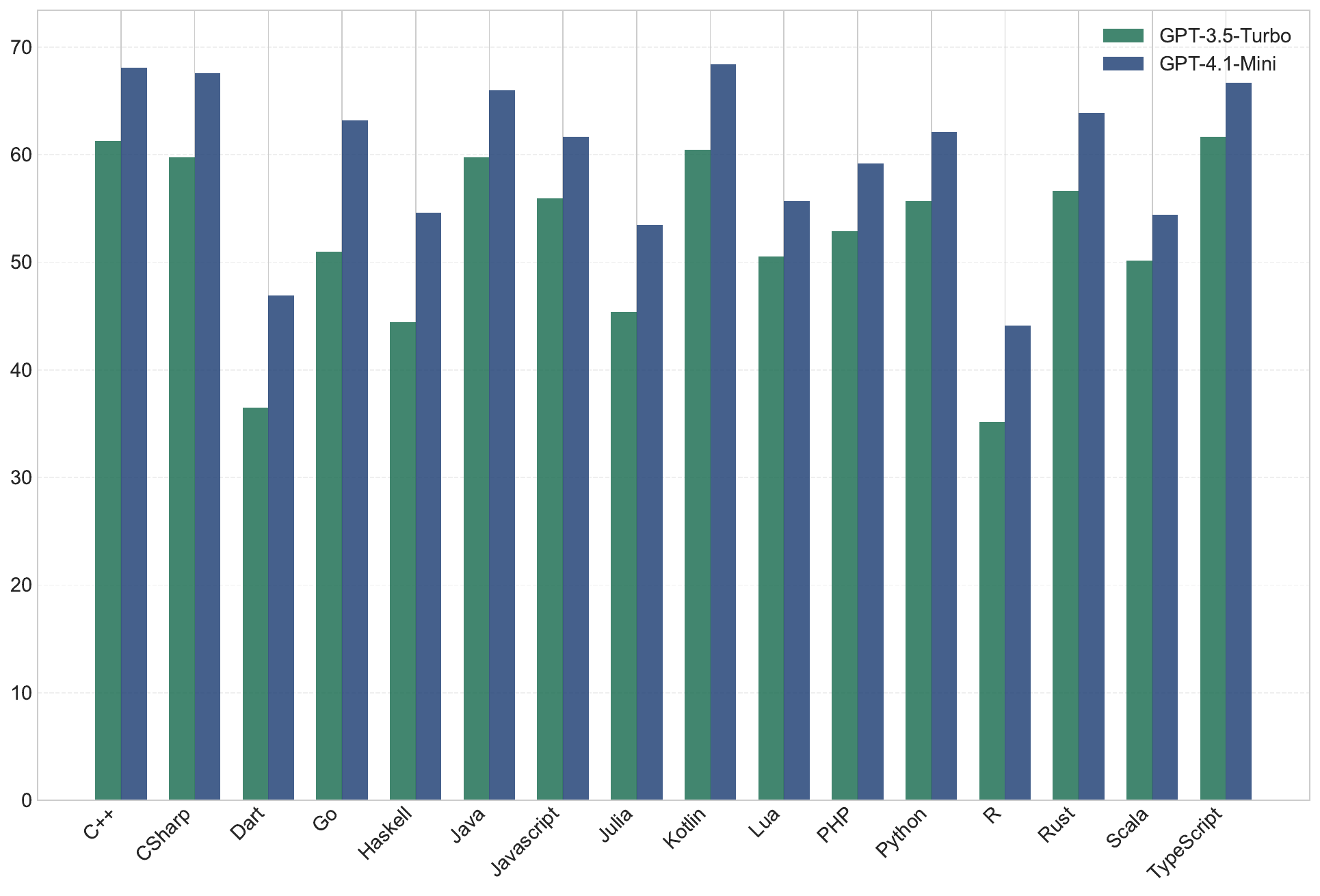} 
         \vspace{-0.5cm}
     \caption{Performance disparities across programming languages on BabelCode-MBPP}
     \label{fig:motivation}
\end{figure}

\textbf{Fine-tuning Dataset}: For language-specific fine-tuning comparisons, we employ the Oxen.ai Rust dataset, containing Rust programming samples generated from translated prompts. This dataset derives from translating 20,000 prompts from the Ace-Code-87k corpus~\cite{zeng2025acecoderacingcoderrl} into Rust using pretrained language models, filtered for successful compilation and unit test execution. The final dataset contains 16,500 prompt–code–unit\_test triplets, partitioned into 15,000 training samples and 1,000 each for testing and validation.\footnote{Dataset available at \url{https://www.oxen.ai/ox/Rust}} For fine-tuning, we convert each training example into a chat-style instance with (i) a Rust-specific system instruction, (ii) a user message containing the problem description and target Rust function signature, and (iii) an assistant message containing the target Rust implementation only.

\subsection{Metrics}
\label{lab:metrics}

We employ a comprehensive evaluation approach centered on functional correctness with detailed error analysis to ensure methodological rigour and practical relevance.

\begin{itemize}[leftmargin=*]
    \item \textbf{Pass@1}~\cite{chen2021evaluating}: Our primary evaluation metric measures the percentage of problems where generated solutions pass all provided test cases. This metric directly captures practical utility and serves as the gold standard in code generation literature, enabling direct comparison with existing work. Pass@1 represents the most realistic deployment scenario, where users evaluate the first generated solution, making it the most practically relevant effectiveness measure.
    
    \item \textbf{Semantic Code Validity Evaluation}: In addition to full success, we track the percentage of tasks for which a generated program passes a non-zero subset of the tests but not all of them. This metric is useful for pipeline analysis because it captures near-miss solutions that are not yet correct but are close enough to benefit from transfer or repair.

    \item \textbf{Error rate.} We also report an aggregated error rate covering failures that prevent full success, including compilation errors and runtime errors. We group these failures together because, in our setting, both types indicate that the generated program could not be validated successfully and therefore require further pipeline intervention.
    
\end{itemize}

The combination of Pass@1 with detailed error categorisation provides both high-level performance indicators and diagnostic insights into specific failure modes, enabling systematic analysis of where and why our approach succeeds or fails while facilitating actionable insights for system improvement.

%% file: sections/results_analysis.tex
\section{Results and Analysis}
\label{sec:evaluation}

\begin{figure*}[!ht]
        \centering
        \includegraphics[width=\linewidth]{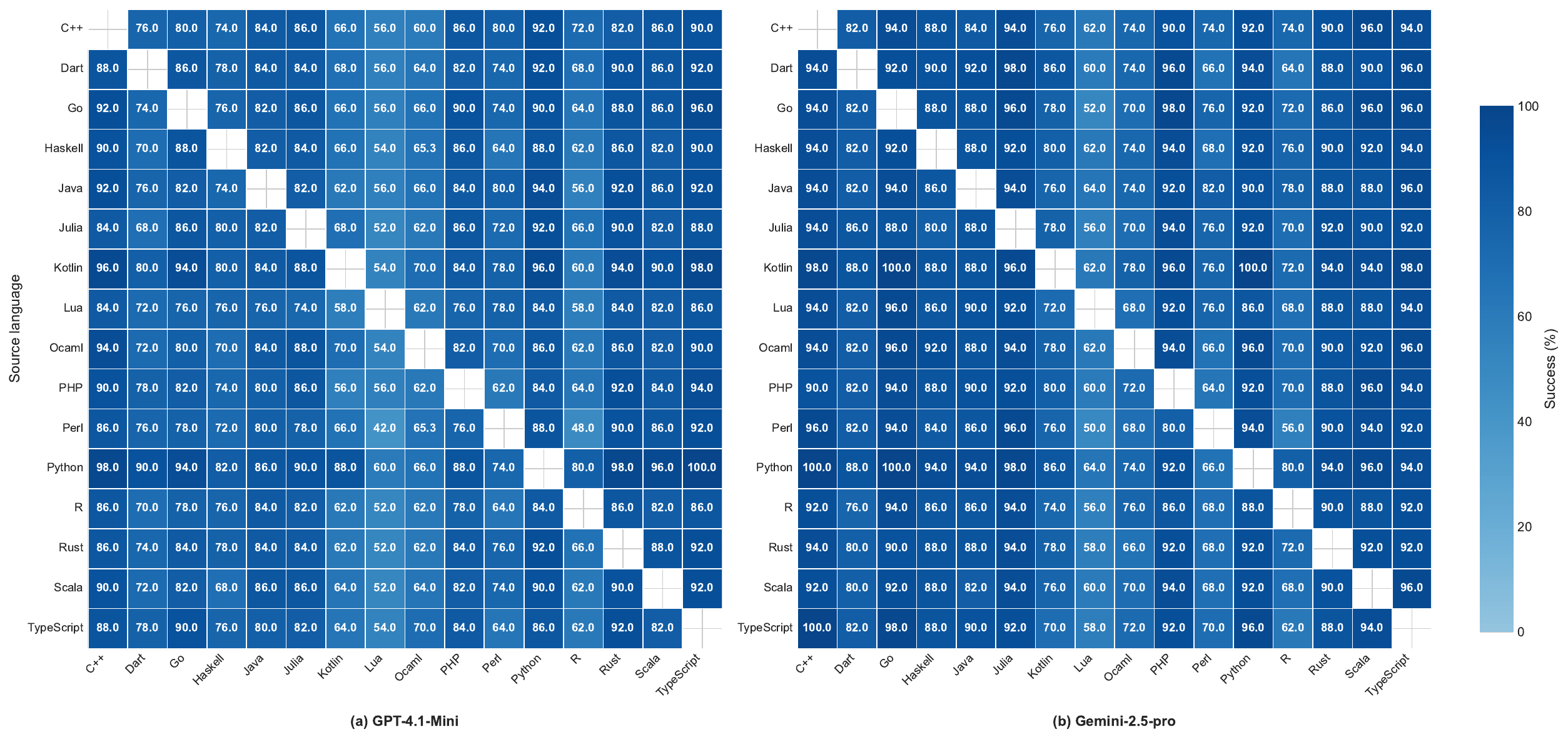} 
        \vspace{-0.5cm}
    \caption{Pairwise code-translation success across programming languages using MultiPL-E. The main figure shows GPT-4.1-Mini and Gemini-2.5-pro; the corresponding DeepSeek-V3.2 heatmap is provided in the appendix.}
    \label{fig:heatmap-translation}
\end{figure*}

\begin{table}[htbp]
\centering
\caption{Full test pass rate (\,\%) by dataset and model.}
\label{tab:full-pass}
\begin{tabular}{@{}lrrr@{}}
\toprule
\textbf{Dataset} & \textbf{GPT-4.1-mini} & \textbf{DeepSeek-V3.2} & \textbf{Gemini 2.5-pro} \\
\midrule
MBPP & 72.6\,\% & 77.5\,\% & 75.9\,\% \\
LeetCode & 57.5\,\% & 51.3\,\% & 65.4\,\% \\
\bottomrule
\end{tabular}
\end{table}

\begin{figure*}[!ht]
        \centering
        \includegraphics[width=\linewidth]{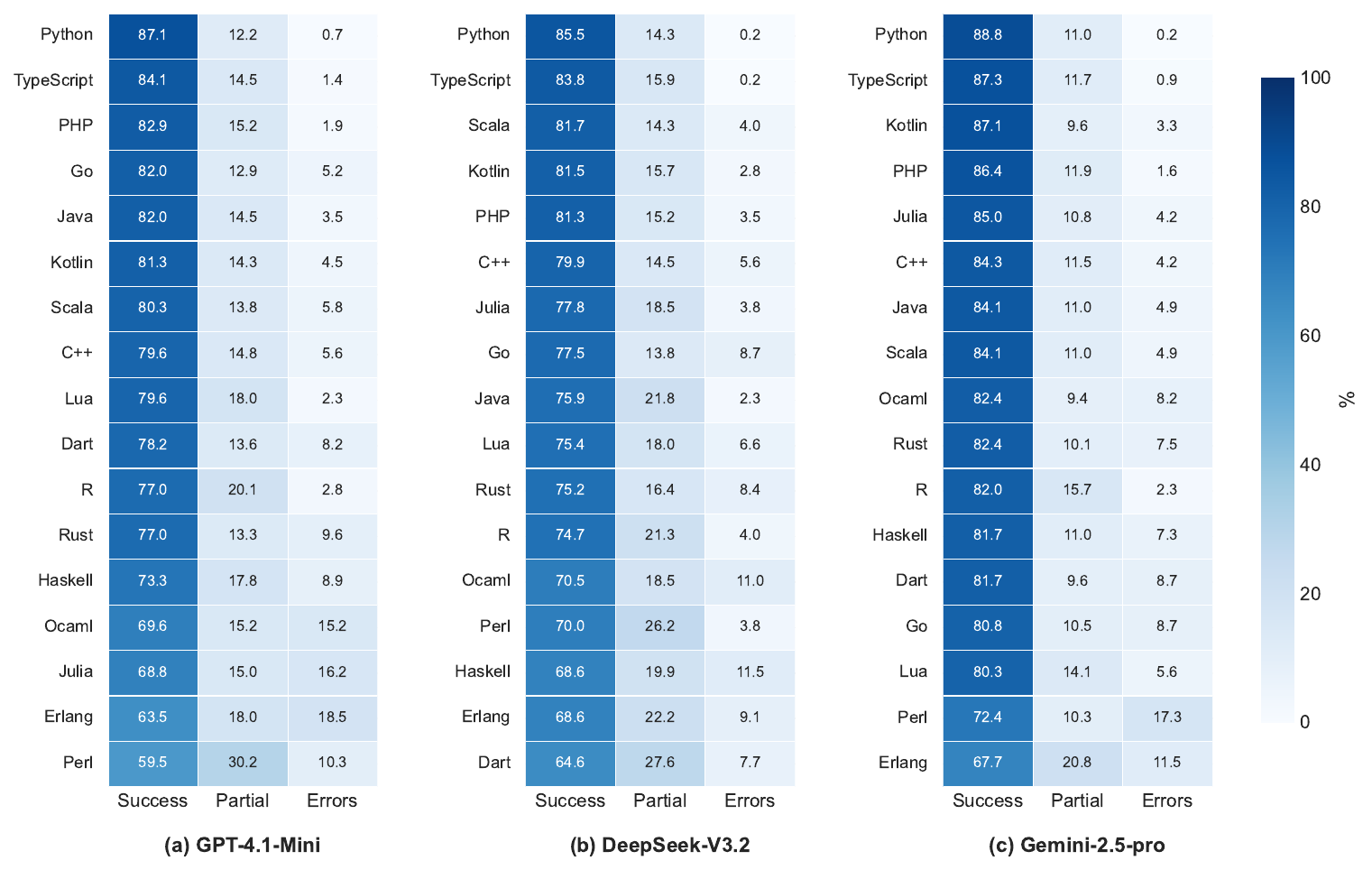} 
        \vspace{-0.5cm}
    \caption{LLM performances in code generation across programming languages (Stage 1, MBPP).}
    \label{fig:stage1}
\end{figure*}

\begin{table}[h]
\caption{Performances of LLMs using translation-based top-3 language selection (Strategic) and LLM-based language selection on MBPP.}
\centering
\renewcommand{\arraystretch}{1.25}
\resizebox{\textwidth}{!}{%
\begin{tabular}{lcccccc}
\hline
\textbf{Languages} & \textbf{DeepSeek-V3.2} & \textbf{DeepSeek-V3.2} & \textbf{GPT4-mini} & \textbf{GPT4-mini} & \textbf{Gemini} & \textbf{Gemini} \\
 & \textbf{(Strategic)} & \textbf{(LLM based)} & \textbf{(Strategic)} & \textbf{(LLM based)} & \textbf{(Strategic)} & \textbf{(LLM based)} \\
\hline
Python & {89.2} & {89.2} & {89.5} & 88.8 & 92.3 & {93.0} \\
TypeScript & {88.1} & 87.8 & {87.4} & 86.9 & 91.6 & {92.0} \\
Dart & {67.7} & {67.7} & - & - & {85.5} & 85.0 \\
PHP & {86.9} & {86.9} & {87.1} & 86.9 & {92.0} & 91.3 \\
Kotlin & {86.2} & 85.9 & - & - & {91.8} & 91.1 \\
Scala & {86.7} & 85.9 & - & - & - & - \\
Perl & 75.9 & {76.6} & 72.8 & {75.4} & 83.4 & {83.8} \\
Rust & 82.0 & {82.2} & 80.8 & {81.0} & 86.2 & {86.9} \\
Ocaml & {79.2} & 78.2 & 77.0 & {77.3} & 87.6 & {87.8} \\
Erlang & 78.7 & {79.2} & {75.2} & 74.9 & 83.8 & {84.1} \\
Haskell & {77.8} & {77.8} & {80.3} & 80.1 & {86.4} & {86.4} \\
Go & - & - & {84.8} & 84.3 & - & - \\
Java & - & - & {84.5} & 84.4 & - & - \\
Julia & - & - & {78.5} & 77.0 & {91.3} & 90.6 \\
\hline
\end{tabular}
}
\label{tab:comparison-mbpp}
\end{table}

\subsection{\textbf{[RQ.1]} How Well Does XL-CoGen Perform in Generating Code Across Programming Languages?}
\label{rq1}
\noindent 
\textbf{Goal.} This research question evaluates the overall effectiveness of XL-CoGen across programming languages and models. We examine direct generation performance, compare it with the full pipeline, and assess whether the same multilingual performance pattern generalises from MBPP to the more challenging LeetCode benchmark.

\begin{figure*}[!ht]
        \centering
        \includegraphics[width=\linewidth]{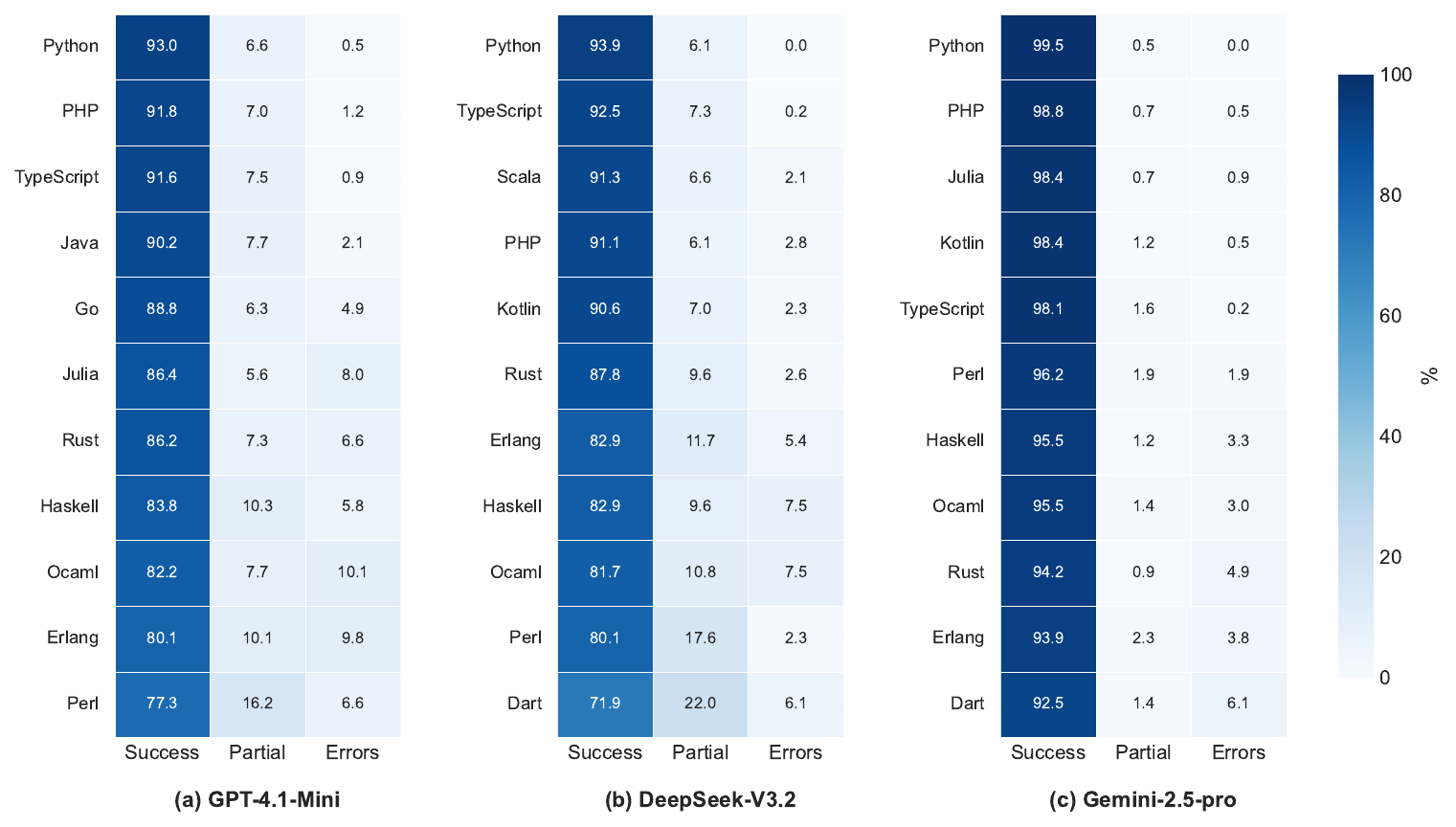} 
        \vspace{-0.5cm}
    \caption{XL-CoGen performances on MBPP.}
    \label{fig:xlcogen}
\end{figure*}

\noindent 
\textbf{Direct-generation baselines on MBPP and LeetCode.} We first report direct-generation performance under the basic benchmark setup as an external reference point for the later pipeline analysis. This baseline is distinct from Stage~1, which already includes harness generation, validation, and correction. Table~\ref{tab:full-pass} shows full-pass rates of 72.6\%--77.5\% on MBPP and 51.3\%--65.4\% on LeetCode across models. This confirms that LeetCode is the harder benchmark, with lower baseline performance under the basic setup. We therefore use MBPP to establish the main multilingual performance pattern and LeetCode to test whether the same pattern holds under greater difficulty.

\noindent
\textbf{Stage 1 performance on MBPP.} We evaluated single-pass (Stage 1) code generation on MBPP across 17 programming languages. The results are shown in Figure~\ref{fig:stage1}. They reveal strongly unbalanced performance: some languages achieve high success rates (e.g., TypeScript and Python) while others remain low (e.g., Perl, Erlang, OCaml). Moreover, language rankings differ across models ---a language that ranks among the top in one model can rank in the middle or lower in another. This variance motivates both a targeted evaluation on a subset of languages that spans the performance spectrum and a data-driven choice of intermediate languages for cross-language transfer.

\noindent
\textbf{Choice of 11 languages per model.} For the remainder of the experiments, we selected 11 languages per model: five high-performing, five low-performing, and one near the middle of the performance distribution. This design ensures coverage of strong and weak generation regimes and allows us to assess how XL-CoGen behaves across the full performance range. The exact sets differ slightly by model (e.g., top-5 and bottom-5 are model-specific), but the 5-5-1 structure is maintained, allowing us to rigorously evaluate the pipeline on both well-supported and challenging languages.

\noindent
\textbf{Translation evaluation and intermediate language selection.} We ran a translation process using MultiPL-E to evaluate code translation from one programming language to another. Figure~\ref{fig:heatmap-translation} reports the resulting pairwise translation success (in \%) for the set of languages evaluated, with separate panels for GPT-4.1-Mini and Gemini-2.5-pro. The heatmap shows that translation success varies strongly by source--target pair: strong source languages remain useful intermediates, but their advantage is not uniform across all targets, and for many targets the best source language is not Python. We then applied a formula that combines (i) MBPP Stage~1 generation performance and (ii) this translation matrix to select, for each target language, the top-3 intermediate languages (ILs) that maximize expected success when transferring solutions. We refer to this data-driven choice as \emph{strategic} selection. Unlike LLM-based selection, which tends to favour languages such as Python regardless of the heatmap, strategic selection selects ILs based on empirical translation success in Figure~\ref{fig:heatmap-translation}. This selection is performed using MBPP and the translation heatmap only; we do not use LeetCode at this stage. On MBPP, we compare pipeline performance when using our strategic top-3 ILs versus when the LLM selects the top-3 intermediate languages (Table~\ref{tab:comparison-mbpp}). In general, we obtain better or equal performance with strategic selection: strategic yields higher pass rates in many language--model pairs (e.g., GPT-4.1-mini Python 89.5\% vs \ 88.8\%, Julia 78.5\% vs \ 77.0\%). Overall, the data-driven choice of ILs yields better or competitive performance than LLM-based selection on MBPP.

\noindent
\textbf{XL-CoGen results on MBPP.} We ran the full XL-CoGen pipeline on MBPP with the strategically chosen ILs. Figure~\ref{fig:xlcogen} summarises the results. Final pass rates remain high on already strong languages and become very strong for the best-performing settings: for example, Python reaches 93.0\% with GPT-4.1-mini, 93.9\% with DeepSeek-V3.2, and 99.5\% with Gemini-2.5-pro, while TypeScript reaches 91.6\%, 92.5\%, and 98.1\%, respectively. For challenging languages, the pipeline delivers substantial recovery across all three models: Rust reaches 86.2\%, 87.8\%, and 94.2\%; Perl 77.3\%, 80.1\%, and 96.2\%; and Erlang 80.1\%, 82.9\%, and 93.9\% for GPT-4.1-mini, DeepSeek-V3.2, and Gemini-2.5-pro, respectively. OCaml and Haskell show the same pattern, reaching 82.2\% and 83.8\% with GPT-4.1-mini, 81.7\% and 82.9\% with DeepSeek-V3.2, and 95.5\% and 95.5\% with Gemini-2.5-pro. Error analysis shows that functional errors are substantially reduced; compilation failures remain more frequent for some languages (e.g., Rust, OCaml). Overall, the pipeline successfully narrows the gap between high- and low-performing languages on MBPP.

\begin{figure*}[!ht]
        \centering
        \includegraphics[width=\linewidth]{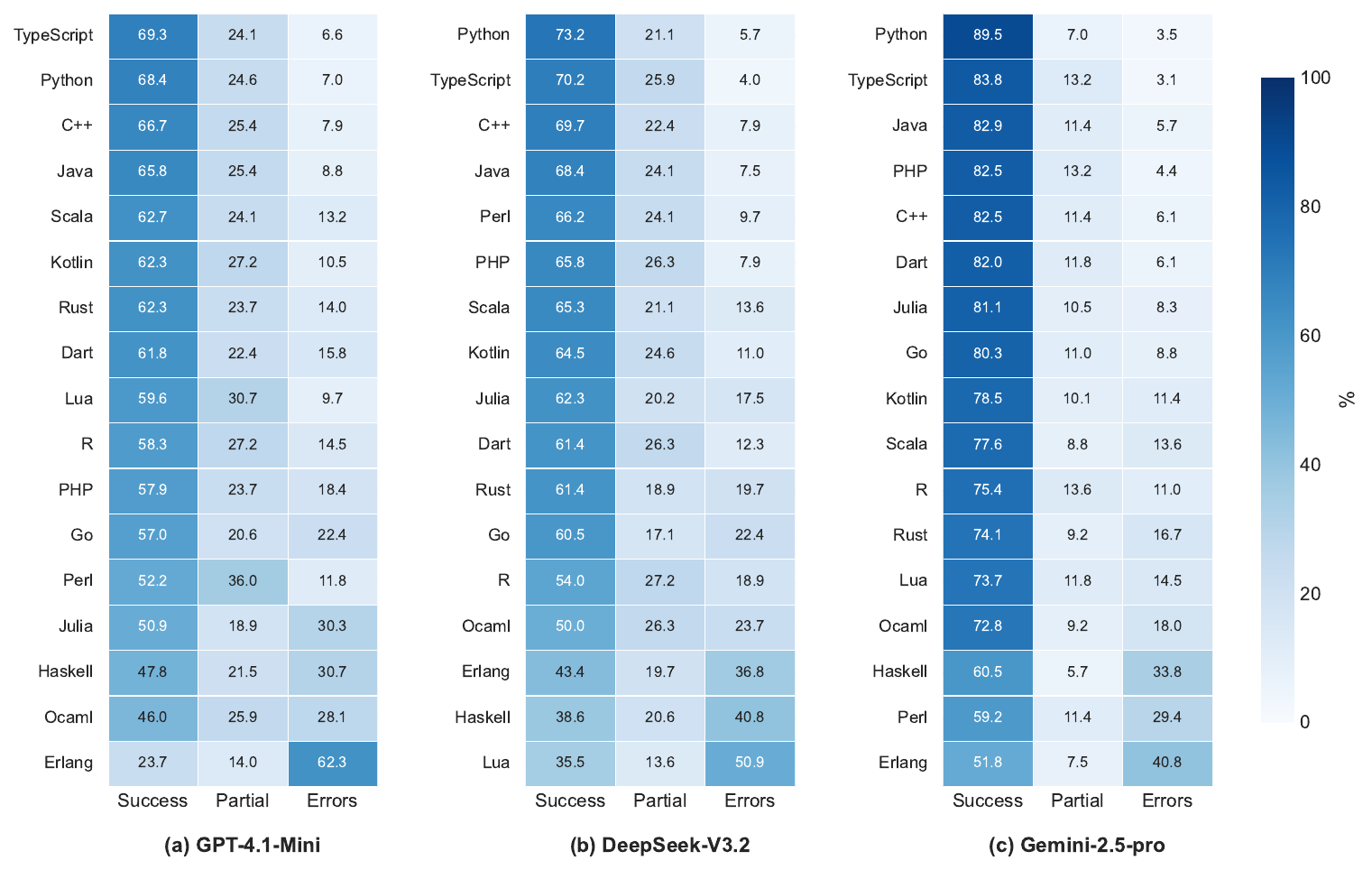} 
        \vspace{-0.5cm}
    \caption{LLM performances in code generation across programming languages (Stage 1, LeetCode).}
    \label{fig:stage1-leetcode}
\end{figure*}

\noindent
\textbf{Stage 1 performance on LeetCode.} We then evaluated single-pass (Stage~1) code generation on the more complex LeetCode benchmark. Figure~\ref{fig:stage1-leetcode} shows the results. As expected from Table~\ref{tab:full-pass}, overall pass rates are lower than on MBPP (e.g., 51.3\%--65.4\% versus 72.6\%--77.5\% by model). The performance distribution across languages is again unbalanced: some languages remain relatively strong while others drop sharply on these harder, less formulaic problems. Model-dependent rankings also persist---the same language can rank differently across models. This confirms that the challenges observed on MBPP (variance across languages and models) carry over to the more complex benchmark, motivating the use of the same pipeline and IL selection for validation.

\noindent
\textbf{Validation of intermediate language selection on LeetCode.} We validated the ILs chosen from MBPP and the translation heatmap by running the same pipeline on LeetCode without changing the intermediate languages. Table~\ref{tab:comparison-leetcode} compares strategic selection with LLM-based selection on LeetCode. The reported differences remain small in some cases, but several entries are higher under strategic selection, including Erlang with DeepSeek-V3.2 (68.4\% vs \ 64.5\%), Ocaml with DeepSeek-V3.2 (68.4\% vs \ 64.0\%), Haskell with DeepSeek-V3.2 (64.5\% vs \ 59.2\%), and Python with GPT-4.1-mini (80.3\% vs \ 76.8\%). There are also cases where LLM-based selection is slightly higher, such as Erlang with GPT-4.1-mini (39.9\% vs.\ 38.2\%). Overall, strategic selection matches or exceeds LLM-based selection in most language-model pairs on LeetCode. This shows that the same ILs derived from MBPP and the translation heatmap remain effective when transferred unchanged to the harder LeetCode benchmark.

\textbf{Generalisation of intermediate-language selection.} A distinctive property of our data-driven selection is that the intermediate-language ranking is computed once, from MBPP Stage~1 performance and the MultiPL-E translation study, and then reused unchanged on LeetCode. This cross-benchmark portability directly addresses a limitation of manual IR-based approaches such as SPEAC, which fix a single intermediate representation per target language and do not address whether such a choice generalises across evaluation contexts. The results in Table~\ref{tab:comparison-leetcode} show that the strategic selection derived from MBPP remains competitive or superior to LLM-based selection on LeetCode in most language--model pairs, suggesting that the merit score captures stable cross-language properties rather than benchmark-specific artefacts.

\begin{figure*}[!ht]
        \centering
        \includegraphics[width=\linewidth]{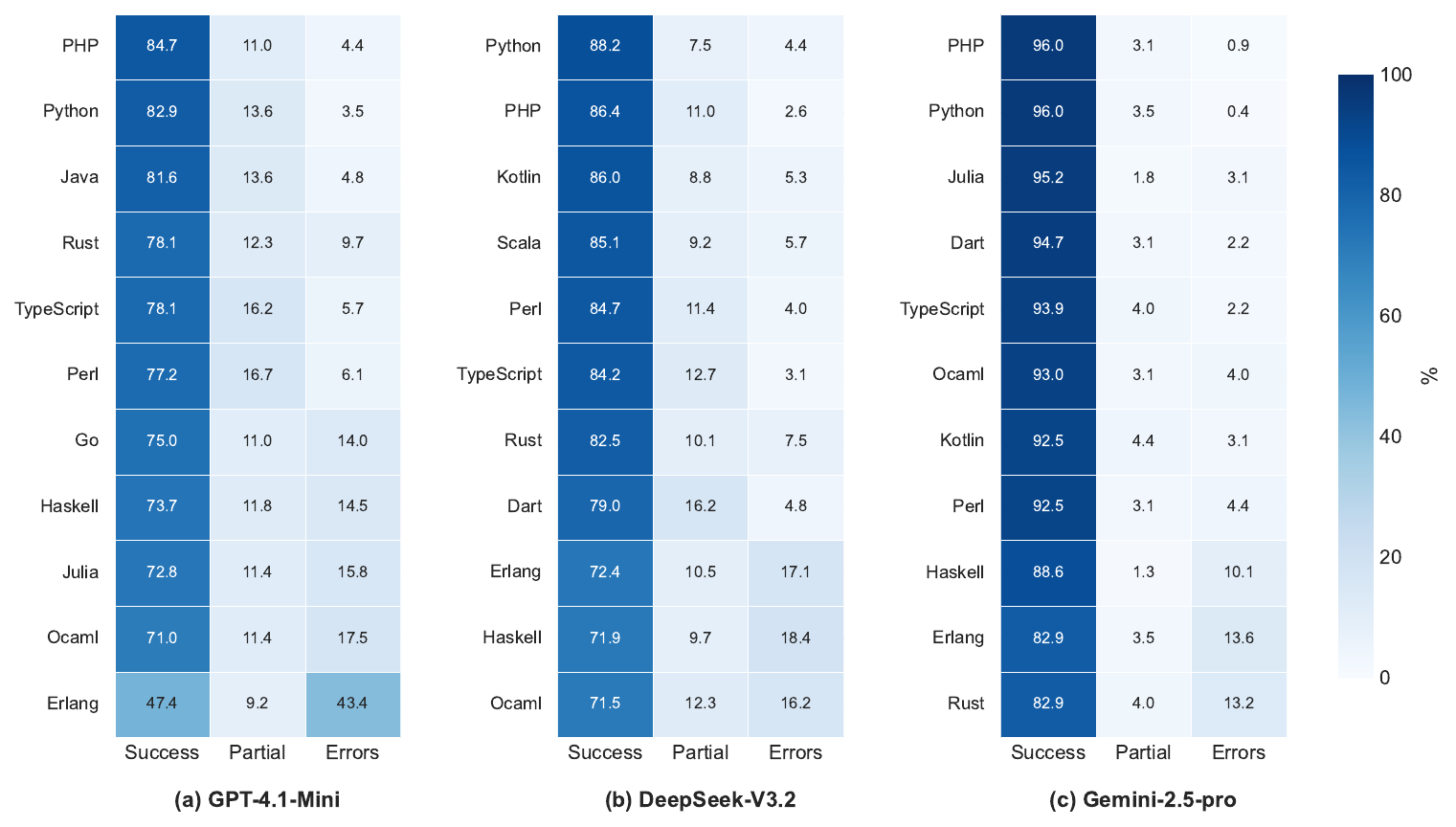} 
        \vspace{-0.5cm}
    \caption{XL-CoGen performances on LeetCode.}
    \label{fig:xlcogen-leetcode}
\end{figure*}

\begin{table}[h]
\caption{Performances of LLMs using translation-based top-3 language selection (Strategic) and LLM-based language selection on LeetCode.}
\centering
\renewcommand{\arraystretch}{1.25}
\resizebox{\textwidth}{!}{%
\begin{tabular}{lcccccc}
\hline
\textbf{Languages} & \textbf{DeepSeek-V3.2} & \textbf{DeepSeek-V3.2} & \textbf{GPT4-mini} & \textbf{GPT4-mini} & \textbf{Gemini} & \textbf{Gemini} \\
 & \textbf{(Strategic)} & \textbf{(LLM based)} & \textbf{(Strategic)} & \textbf{(LLM based)} & \textbf{(Strategic)} & \textbf{(LLM based)} \\
\hline
Python & {85.1} & 84.2 & {80.3} & 76.8 & {94.7} & 93.9 \\
TypeScript & {83.8} & 82.9 & 76.3 & {76.8} & {93.4} & 92.1 \\
Dart & {74.6} & {74.6} & - & - & 91.7 & {92.1} \\
PHP & {83.3} & 82.9 & {78.1} & 75.4 & {94.7} & {94.7} \\
Kotlin & {84.2} & 83.3 & - & - & {92.5} & 90.8 \\
Scala & {82.9} & 82.0 & - & - & - & - \\
Perl & {80.3} & 78.5 & {70.6} & 70.2 & 80.7 & {83.3} \\
Rust & 74.6 & {75.4} & {76.8} & 75.9 & {81.6} & {81.6} \\
Ocaml & {68.4} & 64.0 & 68.4 & {68.9} & {89.0} & 88.6 \\
Erlang & {68.4} & 64.5 & 38.2 & {39.9} & {78.1} & 76.8 \\
Haskell & {64.5} & 59.2 & {67.1} & 66.2 & {81.6} & 81.1 \\
Go & - & - & {73.2} & 68.9 & - & - \\
Java & - & - & {78.1} & 74.6 & - & - \\
Julia & - & - & {68.4} & 67.5 & {91.2} & {91.2} \\
\hline
\end{tabular}}
\label{tab:comparison-leetcode}
\end{table}

\noindent
\textbf{XL-CoGen results on LeetCode.} We ran the full XL-CoGen pipeline on LeetCode using the same strategically chosen ILs. Figure~\ref{fig:xlcogen-leetcode} summarises the results. Because LeetCode is more complex, absolute pass rates are lower than on MBPP; the pipeline nevertheless yields substantial improvements. For high-performing languages, final pass rates on LeetCode are strong (e.g., Python 82.9\%--96.0\%, TypeScript 76.3\%--93.9\% by model). For challenging languages, the pipeline again narrows the gap: Erlang reaches 47.4\%--82.9\%, Haskell 71.9\%--88.6\%, Ocaml 71.0\%--93.0\%, Perl 77.2\%--92.5\%, and Rust 78.1\%--82.9\% depending on model. The magnitude of improvement is comparable to or larger than on MBPP for many languages, especially challenging ones, so the framework generalises from the less complex (MBPP) to the more complex (LeetCode) setting without losing effectiveness.

\textbf{XL-CoGen results on LeetCode.} We ran the full XL-CoGen pipeline on LeetCode using the same strategically chosen ILs. Figure~\ref{fig:xlcogen-leetcode} summarises the results. Because LeetCode is more complex, absolute pass rates are lower than on MBPP, but the pipeline remains effective across models and languages. Final pass rates stay strong on already favourable languages: Python reaches 82.9\% with GPT-4.1-mini, 88.2\% with DeepSeek-V3.2, and 96.0\% with Gemini-2.5-pro, while TypeScript reaches 78.1\%, 84.2\%, and 93.9\%, respectively. The same recovery pattern appears on more challenging languages, where Erlang reaches 47.4\%, 72.4\%, and 82.9\%; Haskell 73.7\%, 71.9\%, and 88.6\%; Ocaml 71.0\%, 71.5\%, and 93.0\%; Perl 77.2\%, 84.7\%, and 92.5\%; and Rust 78.1\%, 82.5\%, and 82.9\% for GPT-4.1-mini, DeepSeek-V3.2, and Gemini-2.5-pro, respectively. Overall, the pipeline continues to narrow the gap between high- and low-performing languages on the harder benchmark, showing that the main multilingual performance pattern established on MBPP carries over to LeetCode.

\find{{\bf [RQ-1] \ding{42} XL-CoGen consistently improves multilingual code generation across programming languages and models, with the strongest gains on initially weak target languages. On MBPP, the full pipeline substantially narrows the gap between strong and weak languages, and the same multilingual recovery pattern carries over to the harder LeetCode benchmark. Strategic intermediate-language selection remains effective on both benchmarks and, when transferred unchanged from MBPP to LeetCode, continues to provide a stronger and more reliable basis for recovery than relying on LLM-based language choice alone.}}

\subsection{\textbf{[RQ.2]} How Does XL-CoGen Compare with Prior Approaches, Including Language-Specific Fine-tuning and Existing Code Generation Tools?}
\label{rq2}
\noindent
\textbf{Goal.} This research question evaluates whether the gains of XL-CoGen come from its multilingual pipeline design rather than from target-language specialisation alone. We compare XL-CoGen against (1) language-specific fine-tuning in a controlled Rust case study and (2) prior code-generation tools on an aligned multilingual subset.

\noindent
\textbf{Fine-tuning setup and process.} We use the Oxen.ai Rust dataset for training: it contains Rust prompt--code--unit\_test triplets derived from the Ace-Code-87k corpus, filtered for successful compilation and test execution (see~\S\ref{lab:datasets}). For \emph{testing}, we use MBPP with the real hand-written test cases, so that evaluation is independent of the fine-tuning data. We first run the \emph{baseline} GPT-4.1-mini on this MBPP test set (Rust instances) to obtain the reference pass rate. We then fine-tune GPT-4.1-mini on the Rust training dataset with three different training sizes---1k, 3k, and 10k samples---to obtain language-specialised variants (FT-1k, FT-3k, FT-10k). The corresponding training configurations are: 1k samples with 3 epochs, batch size 2, and learning-rate multiplier 2; 3k samples with 2 epochs, batch size 4, and learning-rate multiplier 1; and 10k samples with 2 epochs, batch size 4, and learning-rate multiplier 2. Each fine-tuned model is evaluated on the same MBPP test set. We focus on Rust because it is a challenging language in our RQ1 results and because, among the low-performing target languages considered in this study, it offered a suitable curated fine-tuning dataset for a controlled comparison. DeepSeek-V3.2 and Gemini-2.5-pro are not included in this comparison because we did not have comparable controlled fine-tuning support for these models under the same evaluation setup. Performance is measured by Pass@1 (full test pass rate).

\begin{table}[h!]
\centering
\caption{Comparison of fine-tuning approaches vs.\ XL-CoGen on Rust code generation (MBPP test set, Pass@1 \%).}
\begin{tabular}{l|c|c|c|c|c}
\hline
\textbf{Method} & \textbf{Baseline} & \textbf{FT-1k} & \textbf{FT-3k} & \textbf{FT-10k} & \textbf{XL-CoGen} \\
& \textbf{(\%)} & \textbf{(\%)} & \textbf{(\%)} & \textbf{(\%)} & \textbf{(\%)} \\
\hline
Pass@1 & 68.1 & 64.3 & 59.6 & 58 & \textbf{86.2} \\
\hline
\end{tabular}
\label{tab:finetuning}
\end{table}

\noindent
\textbf{Comparison of fine-tuning vs.\ XL-CoGen.} Table~\ref{tab:finetuning} reports the results. The baseline GPT-4.1-mini achieves 68.1\% Pass@1 on the MBPP Rust test set. Fine-tuning degrades performance: FT-1k reaches 64.3\%, FT-3k 59.6\%, and FT-10k 58\%. Increasing the number of training samples thus leads to progressively worse results, suggesting overfitting to the training distribution or loss of general coding ability when over-specialising on Rust. In contrast, XL-CoGen (without any Rust-specific fine-tuning) achieves 86.2\% Pass@1 on the same test set---an 18.1 percentage point improvement over the baseline and a 21.9 point improvement over the best fine-tuned model (FT-1k). For this challenging language, the pipeline design (cross-language transfer and refinement) thus outperforms language-specific fine-tuning and preserves general problem-solving capability without requiring target-language training data.

\begin{table}[htbp]
\centering
\caption{CodeCoT, AgentCoder, LDB, MapCoder and XLCoGen performance on MBPP by language (success rate \%). Bold indicates the best result for each language--model pair.}
\label{tab:mbpp-perf}
\begin{tabular}{llccccc}
\toprule
Model & Method & Haskell & Julia & Python & Rust & Dart \\
\midrule
\multirow{5}{*}{DeepSeek-V3.2} 
 & CodeCoT & 83.8& -- & 83.8& 77.7& 78.5\\
 & AgentCoder & 76.1 & -- & 86.2 & 79.8 & 76.8 \\
 & LDB & 80.6 & -- & 89.7 & 88.0 & 80.3 \\
 & MapCoder & \textbf{85.2} & -- & 92.5 & \textbf{89.4} & \textbf{85.5} \\
 & XLCoGen & 82.9& -- & \textbf{93.9} & 87.8& 71.9\\
\cline{2-7}
\multirow{5}{*}{GPT4.1-mini} 
 & CodeCoT & \textbf{85.0} & 80.8& 81.7& 75.4 & -- \\
 & AgentCoder & 78.7 & 74.5 & 85.7 & 77.0 & -- \\
 & LDB & 82.4 & {80.6} & 88.3 & 83.3 & -- \\
 & MapCoder & 79.9& 82.9& 87.6& 83.1& -- \\
 & XLCoGen & 83.8& \textbf{86.4} & \textbf{93.0} & \textbf{86.2} & -- \\
\cline{2-7}
\multirow{5}{*}{Gemini-2.5-pro} 
 & CodeCoT & 85.2& -- & 90.2& 85.2& 81.7\\
 & AgentCoder & 83.6 & -- & 90.6 & 83.6 & 83.6 \\
 & LDB & {88.3} & -- & 94.2 & 89.4 & 86.4 \\
 & MapCoder & 85.9& -- & 93.2& 87.8& 84.3\\
 & XLCoGen & \textbf{95.5} & -- & \textbf{99.5} & \textbf{94.2} & \textbf{92.5} \\
\bottomrule
\end{tabular}
\end{table}

\textbf{Comparison with existing code-generation tools.} We selected CodeCoT, AgentCoder, LDB, and MapCoder as baselines re-evaluated under aligned multilingual conditions. CodeCoR~\cite{pan2025codecorllmbasedselfreflectivemultiagent} was initially considered but excluded. The repair instructions passed to the Coding Agent, and the candidate pruning mechanism are neither fully specified in the paper nor released in the repository. Our best-effort reproduction yielded pass rates substantially below those reported by Pan et al., suggesting these components are essential to the claimed performance. Including results from an incomplete implementation would risk misrepresenting CodeCoR's capabilities.

Because these baselines were originally developed for predominantly single-language settings, the comparison is necessarily conducted on a shared multilingual subset rather than on the full XL-CoGen evaluation space. To make this comparison fair, we constructed validated executable tests for the target-language instances required by these tools, aligned the resulting test conditions across systems, and removed samples for which cross-tool consistency could not be established. The comparison subset was chosen to span the performance spectrum observed in RQ1: it includes a high-performing language (Python), a challenging mid-range language (Rust), and low-performing languages for which we could manually construct and validate executable target-language tests (Haskell, Julia, and Dart, depending on the model). Because this low-performing subset is not identical across all models, some entries in Table~\ref{tab:mbpp-perf} are left empty. The reported languages are therefore exactly the subset for which CodeCoT, AgentCoder, LDB, MapCoder, and XL-CoGen could all be evaluated under the same validated multilingual conditions.

Under this conservative setup, XL-CoGen obtains the best result in 8 of the 12 reported language--model pairs. It is best in all four reported Gemini-2.5-pro pairs, in three of four GPT-4.1-mini pairs (where CodeCoT remains stronger on Haskell), and in one of four DeepSeek-V3.2 pairs. For DeepSeek-V3.2, XL-CoGen is strongest on Python (93.9\%), while MapCoder remains higher on Haskell, Rust, and Dart. We therefore do not interpret this comparison as evidence of uniform dominance over prior tools. Rather, it shows that when the task is reframed as multilingual generation from a single specification under aligned, validated tests, a pipeline built around cross-language transfer and repair is competitive with strong prior multi-agent systems and often stronger on the reported subset. In this comparison, the gains of XL-CoGen come from explicit multilingual structure---intermediate-language selection, translation, and repair---rather than from language-specific training or code-conditioned test generation.

Two factors help explain the DeepSeek-V3.2 cases where MapCoder outperforms XL-CoGen. First, there is an inherent asymmetry in the evaluation conditions. MapCoder is evaluated against target-language test cases that were manually written and validated for this comparison, whereas XL-CoGen derives its executable evaluation interface automatically from Python-based assertions at runtime, adapting them to each target language without manual intervention. In the comparison, MapCoder benefits from pre-validated tests that eliminate one source of uncertainty entirely. When DeepSeek-V3.2 is already a strong generator in a given target language, this advantage is amplified, as the model only needs to produce correct code without having to jointly construct and validate the evaluation scaffold. This is particularly visible for Dart, where the gap is largest (85.5\% vs 71.9\%), and where adapting Python assertions to a target-language harness introduces more translation complexity. Second, DeepSeek-V3.2 already achieves high standalone generation scores in the affected languages (89.4\% on Rust, 85.2\% on Haskell), which reduces the performance asymmetry that drives XL-CoGen's cross-language transfer advantage. Taken together, these two factors explain why XL-CoGen and MapCoder are complementary rather than strictly comparable, as XL-CoGen is most effective when language asymmetry is high or when only language-agnostic test specifications are available, whereas MapCoder excels when the model is already capable in the target language and reliable target-language tests are provided 
upfront.

\find{{\bf [RQ-2]} \ding{42} The gains of XL-CoGen are better explained by its multilingual pipeline design than by target-language specialisation alone. In the controlled Rust case study, XL-CoGen outperforms both the direct baseline and the fine-tuned variants evaluated under the same MBPP test protocol. In the comparison with prior multi-agent tools on the aligned multilingual subset, XL-CoGen is competitive overall and strongest in most reported language--model pairs. Taken together, these results show that XL-CoGen gains are driven by structured cross-language recovery, not by target-language specialisation alone.}

\subsection{\textbf{[RQ.3]} What is the Impact of Individual Components of XL-CoGen on Its Overall Performance?}
\label{rq3}
\textbf{Goal.} This research question evaluates how each stage of XL-CoGen contributes to multilingual code generation performance, and whether their roles remain the same as task difficulty increases. We report ablation results first on MBPP, then on LeetCode to examine how the relative importance of transfer and repair changes on a harder benchmark.
\noindent\textbf{Experiment.}  
We conduct an ablation study by selectively disabling stages of the pipeline. We define three partial variants: (1) \texttt{Stage 1 only} (harness generation, correction, and code generation in the target language,~\S\ref{sec:stage1}), (2) \texttt{Stage 1 + Stage 2} (transfer without refinement), and (3) \texttt{Stage 1 + Stage 3} (refinement without transfer). Each variant is evaluated on the same problems and languages as in RQ1. We compare these to the full three-stage XL-CoGen system and report per-language success rates at each stage to identify which component contributes most to closing the performance gap.

\subsubsection{Ablation on MBPP.}
Table~\ref{tab:language_success} reports success rates by language and model for Stage 1, Stage 1+2, Stage 1+3, and the full pipeline (All Stages) on MBPP; ~\Cref{tab:language_success} gives a visual summary. Cells marked ``--'' indicate language--model pairs that were not part of this ablation configuration. On MBPP, transfer and repair play more balanced roles than on LeetCode, and the strongest gains remain concentrated on initially weak languages.

\textbf{On MBPP, repair-only is often stronger than transfer-only, but not uniformly.} In many comparable settings, \texttt{Stage 1+3} remains higher than \texttt{Stage 1+2}. This is visible on already moderate or strong settings such as Java with GPT-4.1-Mini (90.2\% vs \ 84.5\%), Rust with DeepSeek-V3.2 (87.1\% vs \ 82.0\%), and PHP with Gemini-2.5-pro (98.4\% vs \ 92.0\%). On MBPP, once Stage 1 already provides a usable target-language candidate, repair often gives the larger single-stage lift.

\textbf{Already on MBPP, Stage 2 overtakes repair on some weak settings.} This pattern appears precisely on the weakest targets for GPT-4.1-Mini and DeepSeek-V3.2: with GPT-4.1-Mini, \texttt{Stage 1+2} exceeds \texttt{Stage 1+3} on Perl (72.8\% vs.\ 67.9\%) and Erlang (75.2\% vs.\ 74.2\%); with DeepSeek-V3.2, it also exceeds \texttt{Stage 1+3} on Erlang (78.7\% vs.\ 78.2\%). These cases show that, already on the easier benchmark, transfer becomes the better recovery mechanism when direct generation is especially weak.

\textbf{The full pipeline is strongest because it combines both mechanisms.} For GPT-4.1-Mini and DeepSeek-V3.2, the full pipeline is stronger than both partial variants in every comparable MBPP cell. For Gemini-2.5-pro, it remains strongest in most cases and stays within 0.5 points of \texttt{Stage 1+3} on Dart, Perl, and Rust. The largest end-to-end gains are again concentrated on weak languages: with GPT-4.1-Mini, Perl improves from 59.5\% to 77.3\%, Erlang from 63.5\% to 80.1\%, and Ocaml from 69.6\% to 82.2\%; with DeepSeek-V3.2, Erlang and Haskell both rise from 68.6\% to 82.9\%; with Gemini-2.5-pro, Erlang rises from 67.7\% to 93.9\% and Perl from 72.4\% to 96.2\%.

\textbf{The full pipeline is strongest because it combines transfer and refinement.} The ablation therefore supports the same interpretation as RQ1: improvement is not driven by a single step in isolation, but by the interaction between cross-language transfer and final correction. On MBPP, this interaction is especially beneficial for low-performing languages, where Stage 1 alone leaves the largest room for recovery.

\begin{table}[h]
\centering
\caption{Ablation: success rate (\,\%) by stage on MBPP.}
\label{tab:language_success}
\renewcommand{\arraystretch}{1.25}
\resizebox{\textwidth}{!}{%
\begin{tabular}{l|cccc|cccc|cccc}
\toprule
 & \multicolumn{4}{c|}{\textbf{GPT-4.1-Mini}} & \multicolumn{4}{c|}{\textbf{DeepSeek-V3.2}} & \multicolumn{4}{c}{\textbf{Gemini-2.5-pro}} \\
\cmidrule(lr){2-5} \cmidrule(lr){6-9} \cmidrule(lr){10-13}
\textbf{Language} & \textbf{Stage 1} & \textbf{Stage 1+2} & \textbf{Stage 1+3} & \textbf{All Stages} & \textbf{Stage 1} & \textbf{Stage 1+2} & \textbf{Stage 1+3} & \textbf{All Stages} & \textbf{Stage 1} & \textbf{Stage 1+2} & \textbf{Stage 1+3} & \textbf{All Stages} \\
\midrule
\rowcolors{1}{}{gray!8}
Dart & -- & -- & -- & -- & 64.6 & 67.7 & 70.3 & \textbf{71.9} & 81.7 & 85.5 & \textbf{93.0} & 92.5 \\
Erlang & 63.5 & 75.2 & 74.2 & \textbf{80.1} & 68.6 & 78.7 & 78.2 & \textbf{82.9} & 67.7 & 83.8 & 92.5 & \textbf{93.9} \\
Go & 82.0 & 84.8 & 87.8 & \textbf{88.8} & -- & -- & -- & -- & -- & -- & -- & -- \\
Haskell & 73.3 & 80.3 & 82.0 & \textbf{83.8} & 68.6 & 77.8 & 79.6 & \textbf{82.9} & 81.7 & 86.4 & 94.8 & \textbf{95.5} \\
Java & 82.0 & 84.5 & \textbf{90.2} & \textbf{90.2} & -- & -- & -- & -- & -- & -- & -- & -- \\
Julia & 68.8 & 78.5 & 83.8 & \textbf{86.4} & -- & -- & -- & -- & 85.0 & 91.3 & 97.0 & \textbf{98.4} \\
Kotlin & -- & -- & -- & -- & 81.5 & 86.2 & 89.0 & \textbf{90.6} & 87.1 & 91.8 & \textbf{98.4} & \textbf{98.4} \\
Ocaml & 69.6 & 77.0 & 79.9 & \textbf{82.2} & 70.5 & 79.2 & 79.4 & \textbf{81.7} & 82.4 & 87.6 & 94.2 & \textbf{95.5} \\
PHP & 82.9 & 87.1 & 89.9 & \textbf{91.8} & 81.3 & 86.9 & 90.2 & \textbf{91.1} & 86.4 & 92.0 & 98.4 & \textbf{98.8} \\
Perl & 59.5 & 72.8 & 67.9 & \textbf{77.3} & 70.0 & 75.9 & 79.9 & \textbf{80.1} & 72.4 & 83.4 & \textbf{96.7} & 96.2 \\
Python & 87.1 & 89.5 & 91.6 & \textbf{93.0} & 85.5 & 89.2 & 93.4 & \textbf{93.9} & 88.8 & 92.3 & 99.3 & \textbf{99.5} \\
Rust & 77.0 & 80.8 & 85.0 & \textbf{86.2} & 75.2 & 82.0 & 87.1 & \textbf{87.8} & 82.4 & 86.2 & \textbf{94.4} & 94.2 \\
Scala & -- & -- & -- & -- & 81.7 & 86.7 & 89.5 & \textbf{91.3} & -- & -- & -- & -- \\
TypeScript & 84.1 & 87.3 & \textbf{91.6} & \textbf{91.6} & 83.8 & 88.1 & 91.8 & \textbf{92.5} & 87.3 & 91.6 & 97.9 & \textbf{98.1} \\
\bottomrule
\end{tabular}
}
\end{table}

\subsubsection{Ablation on LeetCode.}
We run the same ablation on LeetCode to check whether the relative contribution of each stage holds on the more complex benchmark. Table~\ref{tab:language_success_leetcode} reports success rates; Figure~\ref{fig:xlcogen-leetcode} gives a visual summary. Absolute pass rates are lower than on MBPP (Table~\ref{tab:language_success}), but the relationship between the stages becomes sharper.

\textbf{On LeetCode, the balance shifts toward Stage 2, especially for GPT-4.1-Mini and DeepSeek-V3.2.} For GPT-4.1-Mini, \texttt{Stage 1+2} is higher than \texttt{Stage 1+3} across a broad set of languages, including Go (73.2\% vs.\ 66.2\%), Ocaml (68.4\% vs.\ 61.0\%), Rust (76.8\% vs.\ 72.4\%), Python (80.3\% vs.\ 76.3\%), and TypeScript (76.3\% vs.\ 75.0\%). For DeepSeek-V3.2, the same pattern appears on Erlang (68.4\% vs.\ 64.9\%), Ocaml (68.4\% vs.\ 58.8\%), PHP (83.3\% vs.\ 79.4\%), Python (85.1\% vs.\ 79.8\%), and TypeScript (83.8\% vs.\ 79.0\%). On the harder benchmark, transfer therefore more often produces the stronger intermediate candidate for these two models.

\textbf{For Gemini-2.5-pro, the picture is more mixed.} Transfer is still stronger on several important cases, including Erlang (78.1\% vs \ 71.9\%), PHP (94.7\% vs \ 91.2\%), Python (94.7\% vs \ 94.3\%), and TypeScript (93.4\% vs \ 90.3\%). At the same time, \texttt{Stage 1+3} remains higher on Dart, Haskell, Julia, Ocaml, Perl, and Rust. This suggests that on the strongest model, both transfer and repair can already yield strong candidates on LeetCode, even though transfer remains particularly valuable on some of the weaker target languages.

\textbf{The full pipeline is strongest because it combines transfer and repair.} On LeetCode, the full pipeline matches or exceeds both partial variants in every comparable cell, tying the best partial result only on Gemini-2.5-pro for Kotlin (92.5\%). The largest gains again concentrate on weak languages: Erlang rises from 23.7\% to 47.4\% with GPT-4.1-Mini, from 43.4\% to 72.4\% with DeepSeek-V3.2, and from 51.8\% to 82.9\% with Gemini-2.5-pro; Ocaml and Haskell show the same pattern. The harder benchmark, therefore, makes the division of labour clearer: Stage 2 more often supplies the better candidate, and Stage 3 turns that candidate into the best final result.

\begin{table}[h]
        \centering
        \caption{Ablation: success rate (\,\%) by stage on LeetCode.}
        \label{tab:language_success_leetcode}
        \renewcommand{\arraystretch}{1.25}
        \resizebox{\textwidth}{!}{%
        \begin{tabular}{l|cccc|cccc|cccc}
        \toprule
         & \multicolumn{4}{c|}{\textbf{GPT-4.1-Mini}} & \multicolumn{4}{c|}{\textbf{DeepSeek-V3.2}} & \multicolumn{4}{c}{\textbf{Gemini-2.5-pro}} \\
        \cmidrule(lr){2-5} \cmidrule(lr){6-9} \cmidrule(lr){10-13}
        \textbf{Language} & \textbf{Stage 1} & \textbf{Stage 1+2} & \textbf{Stage 1+3} & \textbf{All Stages} & \textbf{Stage 1} & \textbf{Stage 1+2} & \textbf{Stage 1+3} & \textbf{All Stages} & \textbf{Stage 1} & \textbf{Stage 1+2} & \textbf{Stage 1+3} & \textbf{All Stages} \\
        \midrule
        \rowcolors{1}{}{gray!8}
        Dart & -- & -- & -- & -- & 61.4 & 74.6 & 73.2 & \textbf{79.0} & 82.0 & 91.7 & 93.4 & \textbf{94.7} \\
        Erlang & 23.7 & 37.7 & 41.7 & \textbf{47.4} & 43.4 & 68.4 & 64.9 & \textbf{72.4} & 51.8 & 78.1 & 71.9 & \textbf{82.9} \\
        Go & 57.0 & 73.2 & 66.2 & \textbf{75.0} & -- & -- & -- & -- & -- & -- & -- & -- \\
        Haskell & 47.8 & 67.1 & 66.7 & \textbf{73.7} & 38.6 & 64.5 & 61.4 & \textbf{71.9} & 60.5 & 81.6 & 84.2 & \textbf{88.6} \\
        Java & 65.8 & 78.1 & 77.2 & \textbf{81.6} & -- & -- & -- & -- & -- & -- & -- & -- \\
        Julia & 50.9 & 68.4 & 64.5 & \textbf{72.8} & -- & -- & -- & -- & 81.1 & 91.2 & 92.5 & \textbf{95.2} \\
        Kotlin & -- & -- & -- & -- & 64.5 & 84.2 & 74.6 & \textbf{86.0} & 78.5 & \textbf{92.5} & 91.7 & \textbf{92.5} \\
        Ocaml & 46.0 & 68.4 & 61.0 & \textbf{71.0} & 50.0 & 68.4 & 58.8 & \textbf{71.5} & 72.8 & 89.0 & 89.9 & \textbf{93.0} \\
        PHP & 57.9 & 78.1 & 75.0 & \textbf{84.7} & 65.8 & 83.3 & 79.4 & \textbf{86.4} & 82.5 & 94.7 & 91.2 & \textbf{96.0} \\
        Perl & 52.2 & 70.6 & 69.7 & \textbf{77.2} & 66.2 & 80.3 & 81.1 & \textbf{84.7} & 59.2 & 80.7 & 90.3 & \textbf{92.5} \\
        Python & 68.4 & 80.3 & 76.3 & \textbf{82.9} & 73.2 & 85.1 & 79.8 & \textbf{88.2} & 89.5 & 94.7 & 94.3 & \textbf{96.0} \\
        Rust & 62.3 & 76.8 & 72.4 & \textbf{78.1} & 61.4 & 74.6 & 76.8 & \textbf{82.5} & 74.1 & 81.6 & 82.0 & \textbf{82.9} \\
        Scala & -- & -- & -- & -- & 65.3 & 82.9 & 77.2 & \textbf{85.1} & -- & -- & -- & -- \\
        TypeScript & 69.3 & 76.3 & 75.0 & \textbf{78.1} & 70.2 & 83.8 & 79.0 & \textbf{84.2} & 83.8 & 93.4 & 90.3 & \textbf{93.9} \\
        \bottomrule
        \end{tabular}
        }
\end{table}

\find{{\bf [RQ-3]} \ding{42} The gains of XL-CoGen come from how the framework organises recovery across stages, with Stage~2 providing a particularly important contribution. Its value is not translation in isolation, but recovery through validated generation in strategically chosen intermediate languages, which becomes especially important when the initial target-language candidate is weak and the task is more complex. In these settings, Stage~2 more often provides the stronger starting point for GPT-4.1-Mini and DeepSeek-V3.2, while the full pipeline is strongest because Stage~3 then refines that stronger candidate into the best final solution.}

%% file: sections/discussion.tex
\section{Discussion}
\label{sec:discussion}

\subsection{Role of MBPP and LeetCode}

The two benchmarks play different roles in our evaluation, and this distinction is important for interpreting the results. MBPP is our primary benchmark because it offers the most controlled setting for the core question of the paper: can multilingual code generation be improved by combining direct generation, cross-language transfer, and repair under a shared behavioural specification? Its task format matches the inputs required by XL-CoGen, and its relative simplicity makes it well-suited for isolating stage-wise effects, comparing intermediate-language selection strategies, and analysing which parts of the pipeline contribute most to recovery.

LeetCode intervenes for a different reason. It is not used to define the mechanism, but to validate whether the same mechanism remains effective when the problems become harder, less formulaic, and more demanding in terms of reasoning and robustness. This is why MBPP is used to establish the main pattern, while LeetCode is used to test the stability of that pattern under greater difficulty. The fact that the same pipeline and the same strategic intermediate-language selection remain effective on LeetCode strengthens the claim that XL-CoGen is not exploiting only properties of simpler MBPP-style tasks.

This difference in benchmark role also helps explain the ablation results. On MBPP, \texttt{Stage 1+3} is often stronger than \texttt{Stage 1+2}, which suggests that for simpler tasks, a good target-language candidate can often be recovered effectively through repair. On LeetCode, \texttt{Stage 1+2} more often outperforms \texttt{Stage 1+3}, especially on low-performing languages, which shows that as task difficulty increases, obtaining a stronger candidate through transfer becomes more valuable than trying to repair a weak initial solution directly. The full pipeline is strongest because it supports both regimes: repair is enough in many easier cases, whereas transfer becomes crucial when direct generation quality drops.

\subsection{Implications for Multilingual Code Generation}

More broadly, our results suggest that multilingual code generation should not be treated as a collection of independent single-language tasks. Across benchmarks, models, and target languages, XL-CoGen improves performance by structuring recovery around language asymmetry: when direct generation is weak in a target language, it is often more effective to obtain a correct or near-correct candidate in a stronger intermediate language and transfer it than to rely only on additional target-language specialisation.

This point is reinforced by the comparison with language-specific fine-tuning. In our Rust study, XL-CoGen improves over both the unfine-tuned baseline and the strongest fine-tuned setting, indicating that pipeline design can be more effective than additional target-language training for this setting. We do not claim that fine-tuning is ineffective in general. Rather, our results suggest that, for multilingual text-to-code generation from a shared specification, explicit cross-language transfer can be a stronger lever than language-specific adaptation alone.

\subsection{What the Results Suggest About Transfer}

The results also suggest that intermediate-language choice should be treated as an empirical compatibility problem rather than as a heuristic preference for popular high-resource languages. The translation heatmaps and the comparison with LLM-based intermediate-language selection show that the best source language for a given target is often not the most obvious one. This matters because Stage 2 depends on two distinct properties: the ability to generate a correct solution in the source language, and the ability to translate that solution successfully into the target language. Choosing intermediate languages using both signals is therefore more principled than relying on generic model intuition alone.

More broadly, the gains observed on languages such as Erlang, OCaml, Haskell, Rust, and Perl indicate that a major obstacle in multilingual code generation is not only reasoning about the problem itself, but doing so reliably under the syntactic and semantic constraints of the target language. XL-CoGen mitigates this by separating concerns across stages: Stage 1 establishes a trusted evaluation scaffold, Stage 2 searches for stronger candidates through transfer, and Stage 3 performs targeted correction. This staged design helps explain why the framework remains effective across languages with different paradigms and tooling requirements.

\subsection{Threats to Validity and Limitations}

Our study has several limitations that bound the claims we can make.

\textbf{Benchmark scope.} The main evaluation uses MBPP and LeetCode, which are well-suited to our setting because they start from a natural-language specification and an executable test list. However, they remain mostly single-function tasks and do not cover multi-file projects, external dependencies, or longer-horizon software engineering workflows. The results, therefore, support multilingual text-to-code generation in a controlled setting, not full software development in the wild.

\textbf{Dataset construction.} Our multilingual MBPP benchmark is obtained by adapting the original task descriptions to each target language, and our multilingual LeetCode benchmark is derived by extracting a bounded test list from existing assertions and examples. These transformations were designed to preserve the original task meaning and executable behavior, but they may still retain some bias from the source benchmark formulations, especially because both benchmarks were not originally created for multilingual evaluation.

\textbf{Transfer calibration.} The intermediate-language rankings are estimated from a separate MultiPL-E-based translation study using 50 problems and 16 languages. This gives a practical and data-driven basis for Stage 2, but it is still a finite calibration setting. The rankings may change with larger translation samples, additional languages, or other benchmark distributions.

\textbf{Causal interpretation of Stage 2.} Our ablations show that Stage 2 materially contributes to performance, especially on harder tasks and weaker target languages. However, they do not isolate translation effects from every alternative equal-budget recovery strategy in the target language. A stricter causal comparison against additional target-language resampling or repair-only budgets remains future work.

\textbf{Fine-tuning comparison.} Our fine-tuning study focuses on Rust and one available Rust-specific dataset. This is sufficient to show that XL-CoGen can outperform language-specific fine-tuning in a concrete low-performing target language, but it does not justify a universal claim over all languages or all fine-tuning regimes.

\textbf{Data contamination.} As with most evaluations involving contemporary foundation models, we cannot fully rule out benchmark exposure in pretraining. We therefore do not use contamination as an explanation for the results. The central empirical argument of the paper rests instead on relative comparisons under shared models, cross-language performance patterns, stage-wise ablations, and cross-benchmark consistency.

These limitations clearly delineate the scope of this work without affecting its primary conclusion: XL-CoGen stands as a robust approach for multilingual code generation from shared specifications, particularly when direct generation is uneven across target languages. Extending the framework to broader benchmarks and additional model families remains the primary objective for future research.

%% file: sections/conclusion.tex
\section{Conclusion}
\label{sec:conclusion}

We presented XL-CoGen, a three-stage multilingual code-generation pipeline designed to address the uneven reliability of large language models across programming languages. Starting from a single natural-language specification and a shared test list, XL-CoGen combines validated direct generation, transfer through strategically selected intermediate languages, and targeted repair.

Across programming languages, MBPP, LeetCode, and multiple state-of-the-art LLMs, XL-CoGen consistently improves over direct generation, with the largest gains on initially weak target languages. The results show that the pipeline can outperform language-specific fine-tuning in a challenging target language setting, improving over the best fine-tuned Rust baseline by 22 percentage points in our controlled case study. 
They also show that data-driven intermediate-language selection is more reliable than delegating this choice entirely to the LLM, and that Stage 2 and Stage 3 are complementary: repair is often strong on easier tasks, whereas transfer becomes more important on harder problems.

Taken together, these findings support a simple conclusion: multilingual code generation is better addressed by explicitly exploiting cross-language asymmetry than by treating each target language in isolation. Rather than viewing uneven model performance across languages only as a weakness, XL-CoGen shows that it can be turned into a practical source of recovery and improvement.

\textbf{Future Work.} The most important next steps are to extend the evaluation to broader multilingual benchmarks, including settings with more complex project structure, to test additional model families and target languages, and to study stronger causal controls for isolating the contribution of transfer against alternative target-language recovery strategies.